\documentclass[preprint,nofootinbib,showpacs,preprintnumbers,amsmath,amssymb]{revtex4}

\usepackage{graphicx,epsfig}
\usepackage{simplewick}
\usepackage{slashed}
\usepackage[retainorgcmds]{IEEEtrantools}
\usepackage{color}
\usepackage{setspace}

\def\be{\begin{equation}}
\def\ee{\end{equation}}
\def\bea{\begin{eqnarray}}
\def\eea{\end{eqnarray}}
\def\bear{\begin{array}}
\def\eear{\end{array}}
\def\bes{\begin{subequations}}
\def\ees{\end{subequations}}
\newcommand{\MSbar}{\overline{\rm MS}}  
\newcommand{\m}{{\overline m}}

\usepackage{graphics}
\usepackage{graphicx}% Include figure files
\usepackage{dcolumn}% Align table columns on decimal point
\usepackage{bm}% bold math
\usepackage{epsfig}
\usepackage{graphicx}
\usepackage{multirow}
\usepackage{dcolumn}% Align table columns on decimal point
\usepackage{graphicx,epsfig}% 

\newcommand{\MS}{\overline{\rm MS}}
\newcommand{\RS}{\rm RS}

\newcommand{\nn}{\nonumber}
\def\als{\alpha_{s}}

\begin{document}

\preprint{USM-TH-326}

\title{The bottom quark mass from the $\Upsilon(1S)$ system at NNNLO}
\author{C\'esar Ayala$^1$}
 \email{c.ayala86@gmail.com}
\author{Gorazd Cveti\v{c}$^1$} 
 \email{gorazd.cvetic@usm.cl}
 \author{Antonio Pineda$^2$}
 \email{AntonioMiguel.Pineda@uab.es}

\affiliation{$^1$Department of Physics, Universidad T{\'e}cnica Federico
Santa Mar{\'\i}a (UTFSM),  
Casilla 110-V, 
Valpara{\'\i}so, Chile\\
$^2$Grup de F\'{\i}sica Te\`orica, Universitat Aut\`onoma de Barcelona,
E-08193 Bellaterra, Barcelona, Spain}
\date{\today}

\begin{abstract}
We obtain an improved determination of the normalization 
constant of the first infrared renormalon of the pole mass (and the singlet static potential).
For $N_f=3$ it reads $N_m=0.563(26)$.
Charm quark effects in the bottom quark mass determination are carefully investigated.
Finally, we determine the bottom quark mass using the NNNLO perturbative expression for 
the $\Upsilon(1S)$ mass.
We work in the renormalon subtracted scheme, which allows us to control the divergence of the
 perturbation 
series due to pole mass renormalon. 
Our result for the $\MSbar$ mass reads $\m_{b}(\m_{b})=4201(43)$ MeV.
\end{abstract}

\maketitle

\section{Introduction}
\label{intr}

The mass of the heavy quarkonium ground state 
has been computed to increasingly higher order in perturbation theory over the years \cite{Fischler:1977yf,Billoire:1979ih,Schroder:1998vy,Pineda:1997hz,BPSV,Kniehl:2002br,Penin:2002zv,Smirnov:2008pn,Anzai:2009tm,Smirnov:2009fh}, 
presently reaching NNNLO precision, i.e. ${\cal O}(m\als^5)$. The use of effective field theory methods 
\cite{Caswell:1985ui,Bodwin,Pineda:1997bj,Brambilla:1999xf} 
was important for reaching this accuracy. 
One of the main motivations for this ongoing effort is the possibility to obtain accurate determinations of the bottom and (may be)
charm quark masses by equating these theoretical expressions to the experimental values \cite{Pineda:1997hz,
Beneke:1999fe,
Brambilla:2001fw,Pineda:2001zq,Brambilla:2001qk,Penin:2002zv,Lee:2003hh,Contreras:2003zb,Ayala:2012yw}.
This possibility is reinforced by the fact that the renormalon of the pole mass cancels with the 
renormalon of the static potential~\cite{Pineda:1998id,Hoang:1998nz,Beneke:1998rk} making these 
energies mainly perturbative objects, and, therefore, ideal candidates for good determinations of the heavy quark masses. 
Taking advantage of this fact requires the use of the so-called threshold masses (see, for instance, 
\cite{Bigi:1994em,Beneke:1998rk,Hoang:1999zc,Pineda:2001zq}), which explicitly take 
into account the cancellation between the pole mass and the static potential renormalon. 

One of these analysis was made in Ref.~\cite{Pineda:2001zq}. In this reference  
the NNLO expression of the heavy quarkonium ground state mass was used, as well as some partial NNNLO effects (those obtained from 
the large-$\beta_0$ approximation \cite{Kiyo:2000fr,Hoang:2000fm}, as well as the leading ${\cal O}(m\als^5)$ logarithms \cite{BPSV}) 
to obtain an accurate determination of the 
bottom mass. The charm quark was considered to be active and its mass approximated to zero. 
The threshold mass used was the so-called renormalon subtracted (RS) mass. This mass is defined such that the leading renormalon 
of the pole mass is explicitly subtracted. 
Therefore, it requires the knowledge of
the normalization of the renormalon, which was also approximately computed in that reference. 
 
There is a series of developments that motivate updating such analysis. One obvious improvement would be  
the incorporation of the charm quark mass effects. For the case of the heavy quarkonium mass (versus the pole mass) 
it was concluded that the charm quark decoupled and it was a good approximation to consider the theory with only three active massless flavours \cite{Brambilla:2001qk}. In this paper we argue that one should also use this approximation for the relation between the pole and the $\MS$ mass, producing much smaller shifts than if working with four active flavours. Therefore, the calculation should be redone accordingly. 

It is also possible to improve the determination of the normalization of the pole mass (and the static potential) leading renormalon, $N_m$. 
On the one hand the existence of the three-loop expression of the static potential allows the determination of $N_m$ 
to one order higher in the corresponding expansion. On the other hand, recent analysis in lattice simulations \cite{Bali:2013pla} 
suggest that the direct determination of $N_m$ from the last known coefficients of the perturbative series may actually produce more accurate results 
than previous estimates, which were obtained using the Borel transform of the perturbative series as their key quantity. 
Such improved value would have an immediate impact in heavy quark physics in general, and in the 
determination of the heavy quark mass from the heavy quarkonium spectrum in particular. 

With respect to the latter, the complete NNNLO correction to the perturbative expression of the 
$\Upsilon(1S)$ mass is now known. By including the complete NNNLO expression we can study this term without 
scheme ambiguity and assess its impact. Even more important, at this order ultrasoft effects appear for the first time. 
There is the worry that physics at the ultrasoft scale can not be computed in perturbation theory. The reason is that the natural 
scale associated to those degrees of freedom is of order $m\als^2$, which, up to numerical factors, is 
a low scale. Yet, there have been some analysis where the ultrasoft scale has been treated in perturbation theory. 
For instance, in Ref~\cite{Brambilla:2010pp} the perturbative expression of the static potential (which includes ultrasoft effects) was 
compared with lattice simulations.  Also in Refs.~\cite{Brambilla:2001fw,Brambilla:2001qk}
it has been argued that the nonperturbative effects are small for the low states of heavy quarkonium. 
On the other hand two existing analyses that incorporate the NNNLO expression yield bigger values 
of $m_b$ \cite{Penin:2002zv,Ayala:2012yw}.\footnote{Ref.~\cite{Penin:2002zv} uses an estimate for the three-loop static potential.} 
In our analysis we would like to quantify the real impact of these corrections, as 
it is important to know what pure perturbation theory has to say before asking for non-perturbative corrections.
Finally, the existence of the complete NNNLO result allows to compare with the large-$\beta_0$ estimate of the NNNLO correction, and see how reliable such approximation is.

In Sec.~\ref{sec:charm} we study the corrections to the pole mass and the static potential due to the charm quark. 
In Sec.~\ref{TB} we present the calculation of the normalization
constant (residue parameter) $N_m$ of the leading infrared renormalon:
in Subsec.~\ref{sub:p} from the static quark potential $V(r)$ and 
in Subsec.~\ref{sub:hm} from the ratio $m_q/\m_q$. In addition, in
Subsec.~\ref{Sec:Nm} we elaborate over these determinations and 
obtain an improved estimate of $N_m$ and the coefficients
$r_3$ and $r_4$ of the perturbative expansion in $\als$ of the mass ratio $m_q/\m_q$. In Sec.~\ref{Sec:mbot}
we extract the bottom quark mass from the energy of the
quarkonium ground state in the RS and RS' scheme defined in Ref.~\cite{Pineda:2001zq} and later in the text. In the conclusions
we summarize the results obtained.  

\section{Charm effects in the pole mass and static potential}
\label{sec:charm}

In this section we present the perturbation 
expansions of the pole mass and static potential with special emphasis 
as to how to incorporate charm quark effects.

\subsection{Charm quark effects in the pole mass}
\label{sec:charmmOS}

In this subsection we assume that we have $N_l$ massless quarks, one active massive quark with mass $m_c$, and a (non-active) heavy quark with mass $m_b$ (such that $m_b > m_c$). Therefore, we have a total of $N_f=N_l+1$ active quarks. 
This is the situation relevant for the bottom quark (where $N_f=4$ and $N_l=3$). 

The pole mass $m_b$ and the $\MS$ mass $\m_b \equiv\m_b(\mu=\m_b)$ of the quark $b$ 
are related by the following equality
\bea
m_b&=&\m_b
\left[
1 + R_0 \left(a_{+}(\m_b) + r_1(N_f) a_{+}^2(\m_b) + 
r_2(N_f) a_{+}^3(\m_b)
+ r_3(N_f) a_{+}^4(\m_b)\right)  + {\cal O}(a_+^5) 
\right]
\nonumber\\
&&+\delta m_c^{(+)}
\ ,
\label{mqbmq}
\eea
where the coefficients $r_n$ have been evaluated with $N_f$ active massless quarks. The "+" stands for the fact that $a$ is also evaluated with $N_f$ (massless) 
active quarks: $a_{+}(\mu) = a(\mu;N_f)\equiv \als(\mu;N_f)/\pi$.  
The coefficients $R_0$, $r_1$, and $r_2$ were obtained in Refs.~\cite{Tarrach:1980up}, \cite{Gray:1990yh},
\cite{Chetyrkin:1999ys,Melnikov:2000qh}, respectively:
\be
R_0 = \frac{4}{3} \ , 
\quad R_0r_1(N_f) = 6.248 \beta_0 - 3.739 \ ,
\quad 
R_0r_2(N_f)  =   23.497 \beta_0^2 + 6.248 \beta_1 
+ 1.019 \beta_0 - 29.94 \ .
\label{rs}
\ee
The value of $r_3$ is unknown (except for the $N_l^3$ \cite{Beneke:1994qe} and $N_l^2$ \cite{Lee:2013sx} dependence). Therefore, the value of $r_3$ will be estimated, see Table \ref{tab:r3} in Sec.~\ref{Sec:Nm}. 
Here, $\beta_0 = (11 - 2 N_f/3)/4$ and $\beta_1 = (102 - 38 N_f/3)/16$ are the 
first two coefficients of the renormalization group equation
of $a$
\be
\frac{d a(Q)}{d \ln Q^2} = - \beta_0 a^2(Q)
\left(  1 + c_1  a(Q) + c_2 a^2(Q) + 
c_3 a^3(Q) +  \cdots \right) \ ,
\label{RGE}
\ee
where we use the notation $c_j \equiv \beta_j/\beta_0$ for $j \geq 1$. $\beta_3$
was computed in Refs.~\cite{vanRitbergen:1997va,Czakon:2004bu}.
Specific values of the coefficients $r_j$ are:
$r_1(N_f=3)=7.739$ and $r_2(N_f=3)=87.224$; and $r_1(N_f=4)=6.958$ and $r_2(N_f=4)=70.659$.

The sum in Eq.~(\ref{mqbmq}) can be reexpressed in
terms of $a_+(\mu)$ at an arbitrary renormalization scale 
$\mu$: 
\begin{subequations}
\label{Sm}
\be
m_b=\m_b\left( 1+ S(N_f)
\right)+\delta m^{(+)}_c\ ,
\ee
where ($r^{(+)}_i(\mu)\equiv r_i(\mu;N_f)$)
\begin{eqnarray}
\label{Smexp}
S(N_f)&=&\frac{4}{3} a_+(\mu)
\left[ 1 + r^{(+)}_1(\mu)  a_+(\mu) + r^{(+)}_2(\mu) a_+^2(\mu) + r^{(+)}_3(\mu) a_+^3(\mu) + {\cal O}(a_+^4) \right] 
\\
r_1(\mu; N_f) & = & r_1(N_f) + \beta_0 L_m(\mu) 
\ ,
\label{r1mu} 
\\
r_2(\mu; N_f) & = & r_2(N_f) + \left( 2 \beta_0  L_m(\mu) r_1  + \beta_0^2 L_m^2(\mu) 
\right) + \beta_1 L_m(\mu) \ , 
\label{r2mu}
\\
r_3(\mu; N_f)  & = & r_3(N_f) + \left( 3 \beta_0 L_m(\mu) r_2 + 3 \beta_0^2 L_m^2(\mu) r_1 + \beta_0^3 L_m^3(\mu) \right)  
\nonumber\\ &&
+ \beta_1 \left( 2 L_m(\mu) r_1 + \frac{5}{2} \beta_0 L_m^2(\mu) \right)
+ \beta_2  L_m(\mu) \ ,
\label{r3mu}
\end{eqnarray}
\end{subequations}
$L_m(\mu) = \ln(\mu^2/\m_b^2)$, and we maintain,
for simplicity, the notation $r_j \equiv r_j(\m_b)$.

Finite-mass charm effects are incorporated in 
\be
\delta m_c^{(+)}=\delta m_{(c,+)}^{(1)}a_{+}^{2}(\m_b)+\delta m_{(c,+)}^{(2)}a_{+}^{3}(\m_b)+{\cal O}(a_{+}^{4})
\,,
\ee 
which vanishes in the $m_c \rightarrow 0$ limit. The first term of the series is known \cite{Gray:1990yh} and reads
\be
\label{deltamc1}
\delta m_{(c,+)}^{(1)}=\frac{4}{3}\m_b\Delta[\m_c/\m_b],
\ee
where (see also \cite{Brambilla:2001qk}) 
\bea
\Delta[r] &=& \frac{1}{4} {\Big [} \ln^2 r + \frac{\pi^2}{6}
- \left( \ln r + \frac{3}{2} \right) r^2 
\nonumber\\
&&+ (1+r) (1 + r^3) \left( {\rm Li}_2(-r) - \frac{1}{2} \ln^2 r +
\log r \log(1+r) + \frac{\pi^2}{6} \right)
\nonumber\\
&& + (1-r) (1 - r^3) \left( {\rm Li}_2(r) - \frac{1}{2} \ln^2 r +
\ln r \ln(1-r) - \frac{\pi^2}{3} \right)
{\Big ]} \ .
\label{Delta}
\eea
%\bea
%\nn
%\Delta[r]&=&-\sum _{n=3}^{\infty } r^{2 n} \left(F'(n)+2 F(n) \log (r)\right)+r^4 \left(-\left(\frac{\log ^2(r)}{4}-\frac{13 \log (r)}{24}+\frac{\pi^2}{24}+\frac{151}{288}\right)\right)
%\\
%   &&+\frac{\pi ^2 r^3}{8}-\frac{3 r^2}{4}+\frac{\pi ^2 r}{8}
%   \,,
%   \\
%F[n]&=&\frac{3 (n-1)}{4 (n-2) n (2 n-3) (2 n-1)}
%\,.
%\eea
The exact expression of $\delta m_{(c,+)}^{(2)}$ was obtained in Ref.~\cite{Bekavac:2007tk} and it will be considered later. 
On the other hand ${\cal O}(a^4_+)$ terms or higher are unknown.

It has been noticed in Ref.~\cite{Hoang:1999us} that $\delta m_{(c,+)}^{(1)}$ is mainly determined by the infrared 
behaviour of the loop integral, which is saturated to a large extent by virtualities of order $\sim m_c$.
In this approximation we have
\be
\label{dm1lineal}
\delta m_{(c,+)}^{(1)} \simeq \m_c\frac{\pi ^2}{6}=2.08907\;{\rm MeV}
\,,
\ee  
for $\m_c=1.27$ GeV (to be compared with the exact result $\delta m_{(c,+)}^{(1)}=1.8058\;{\rm MeV}$ with $\m_b=4.2$ GeV).
This shows that already at this order $\delta m_c^{(+)}$ is dominated by the infrared behaviour of the loop integral. We expect that this will be even more so at higher loops. On the other hand the infrared behaviour of $\delta m_c^{(+)}$ 
can be related with the infrared behaviour of the static potential. For the static potential the charm mass dependence is known with two loop accuracy~\cite{Melles:2000dq}. In Ref.~\cite{Hoang:2000fm} this observation was used to obtain the infrared behaviour of $\delta m_{(c,+)}^{(2)}$, ie. the linear behavior of $\delta m_{(c,+)}^{(2)}$:
\bea
\delta m_{(c,+)}^{(2)}&\simeq&\frac{\m_c\pi^2}{3} \left(\beta_0 \left(\ln
   \left(\frac{\m_b^2}{\m_c^2}\right)+\frac{14}{3}-4 \ln (2)\right)+\frac{19 (b_1
   b_2+f_1 f_2)}{3 \pi }-\frac{59}{45}-\frac{2}{3} \ln (2)\right)
\nn
 \\
 &&+\frac{2}{9}\m_c\pi^2+{\cal O}(\m_c^2)
    \,,
   \label{deltamc2bis}
\eea
where $b_2 = 1.12$; $f_2 = 0.47$; $f_1 = \ln(A/b_2)/\ln(f_2/b_2)$; $b_1 = \ln(A/f_2)/\ln(b_2/f_2)$ and $\ln(A)=\frac{13 \zeta (3)}{19}+\frac{161}{228}-\ln (2)$. The coefficients $b_2$, $f_2$ were obtained from an approximate numerical fit to ~\cite{Melles:2000dq}. 
The last term in Eq. (\ref{deltamc2bis}) comes from the fact that we are using the $\MS$ charm mass (otherwise it could be absorbed in Eq.~(\ref{dm1lineal})). In any case it is small compared with the rest of the coefficient. Eq.~(\ref{deltamc2bis}) is then approximated by
\bea
\nn
\delta m_{(c,+)}^{(2,\rm app)}&\simeq&
\m_c\left(21.277 - 16.998
\ln \left(\frac{\m_c}{\m_b}\right) + N_l(1.097 \ln \left(\frac{\m_c}{\m_b}\right)-1.039 ) + \frac{2}{9} \pi^2\right)
\\
&=&
46.6725 \;{\rm MeV}
   \,.
   \label{deltamc2}
\eea
The exact analytic expression of $\delta m_{(c,+)}^{(2)}$ is extremely lengthy. 
An accurate approximated numerical form can be found in Ref.~\cite{Bekavac:2007tk}, which is enough for our purposes. For our values of 
the bottom and charm masses ($\m_b=4.2$ GeV and $\m_c=1.27$ GeV) it reads
\be
\label{exact}
\delta m_{(c,+)}^{(2)}=48.6793\;{\rm MeV}
\,,
\ee
and its linear approximation reads
\bea
\nn
\delta m_{(c,+)}^{(2,\rm lin)}
&=&
\m_c\left(19.996 - 16.998
\ln \left(\frac{\m_c}{\m_b}\right) + N_l( 1.097 \ln \left(\frac{\m_c}{\m_b}\right)-1.039) + \frac{2}{9} \pi^2\right)
\\
&=&
45.0454\;{\rm MeV}
\,.
\label{linearexact}
\eea
The difference between $\delta m_{(c,+)}^{(2,\rm app)}$ and $\delta m_{(c,+)}^{(2,\rm lin)}$ is due to the approximations involved 
in obtaining Eq.~(\ref{deltamc2}) 
(see the discussion in Ref.~\cite{Bekavac:2007tk}). Therefore, we take Eq.~(\ref{linearexact}) as the exact expression for the linear approximation. 
We observe that the linear approximation represents a quite good approximation of the exact result. 

We can now compare the size of $\delta m_{(c,+)}^{(1)}a_+^2(\m_b) = 9.3$ MeV versus $\delta m_{(c,+)}^{(2)}a_+^3(\m_b) = 18.1$ MeV
($\delta m_{(c,+)}^{(2,\rm lin)}a_+^3(\m_b) = 16.8$ MeV yields a similar number). We observe a bad convergent series (this is also so if we choose different renormalization scales). 
The reason for this bad behaviour is the following. 
In principle, it may seem natural to work with $N_f$ active flavours in Eq.~(\ref{mqbmq}), 
since the natural scale is $m_b \gg m_c$. Nevertheless, as it has been discussed in Ref.~\cite{Ball:1995ni},  at high orders in perturbation theory the 
charm quark decouples. The reason is that at order $n$, the natural scale of the loop integral is $m e^{-n}$, which for $n$ large enough becomes smaller than $m_c$. Therefore, the charm mass 
acts as an infrared cutoff killing the low energy contributions to the integral of the fourth flavour
that would otherwise produce the factorial behaviour. Thus, working with $N_f$ active flavours produces spurious contributions that deteriorate the convergence of $\delta m_c^{(+)}$. This problem can be solved by decoupling the charm quark by expanding 
$a_+$ in powers of $a_-\equiv a(\mu;N_l)$ (see App. \ref{app3} for details). The relation between the pole and the $\MS$ mass now reads 
\be
m_b=\m_b\left( 1+ S(N_l)
\right)+\delta m_c\ ,
\label{SmNl0}
\ee
where ($r^{(-)}_i(\mu)\equiv r_i(\mu;N_l)$)
\be
\label{SmNl}
S(N_l)=\frac{4}{3} a_-(\mu)
\left[ 1 + r^{(-)}_1(\mu)  a_-(\mu) + r^{(-)}_2(\mu) a_-^2(\mu) + r^{(-)}_3(\mu) a_-^3(\mu) + {\cal O}(a_-^4) \right] 
\ee
and we have absorbed the effects of the decoupling of $S$ in $\delta m_c$, which now reads
\be
\delta m_c=\left[\delta m_{(c,+)}^{(1)}+\delta m_{(c,\rm dec.)}^{(1)}
\right]a_{-}^{2}(\m_b)+\left[\delta m_{(c,+)}^{(2)}+\delta m_{(c,dec.)}^{(2)}
\right]a_{-}^{3}(\m_b)+{\cal O}(a_-^4)
\,,
\label{delmc}
\ee
where $\delta m_{(c,\rm dec.)}^{(i)}$ are generated by this decoupling and read
\bea
 \delta  m_{(c,\rm dec.)}^{(1)}=
\frac{2}{9}{\m_b}\left({\rm ln}\left(\frac{\m_b^2}{\m_c^2}\right)-\frac{71}{32}-\frac{\pi^2}{4} \right)
\,,
%&&
%\frac{1}{3} \left(\frac{2}{3} \log \left(
%  \m_b^2/\m_c^2\right)
%   \right.
%   \\
%   &&
%    \left.
%   +3 \left(-\frac{\zeta (3)}{6}+\frac{\pi ^2}{9}+\frac{2195}{288}+\frac{1}{9} \pi ^2 \log
%   (2)\right)\right)
%   \\
%   &&
% +\frac{\zeta (3)}{6}-\frac{\pi ^2}{6}-\frac{779}{96}-\frac{1}{9} \pi ^2 \log (2)
\eea
\begin{eqnarray}
  \delta m_{(c, \rm dec)}^{(2)}  &=&{\m_b}\left(-\frac{2293}{243}-\frac{809}{648}\pi^2+\frac{61}{1944}\pi^4-\frac{11}{81}\pi^2{\rm ln}(2)
  +\frac{2}{81}\pi^2\ln^2(2)+\frac{{\rm ln}^4(2)}{81}
  \right.
\nonumber\\
&&
\left.
+\frac{{3107}}{864}{\rm ln}\left(\frac{\bar{m}_b^2}{\bar{m}_c^2}\right)+\frac{1}{27}\pi^2{\rm ln}\left(\frac{\bar{m}_b^2}{\bar{m}_c^2}\right)+\frac{1}{27}\pi^2{\rm ln}(2){\rm ln}\left(\frac{\bar{m}_b^2}{\bar{m}_c^2}\right)+\frac{1}{27}{\rm ln}^2\left(\frac{\bar{m}_b^2}{\bar{m}_c^2}\right)
 \right.
\nonumber\\
&&
\left.
+\frac{8}{27}{\rm Li}_4\left(\frac{1}{2}\right)-\frac{527}{216}\zeta(3)-\frac{1}{18}\zeta(3){\rm ln}\left(\frac{\bar{m}_b^2}{\bar{m}_c^2}\right) \right)
+\frac{1}{3}{\rm ln}\left(\frac{\bar{m}_b^2}{\bar{m}_c^2} \right)\delta m_{(c,+)}^{(1)}
\,.
    \end{eqnarray}

\begin{figure}[htb] %\unitlength=1mm
%\centering\epsfig{file=mb(mu).eps,width=8.cm}
\centering\includegraphics[width=120mm]{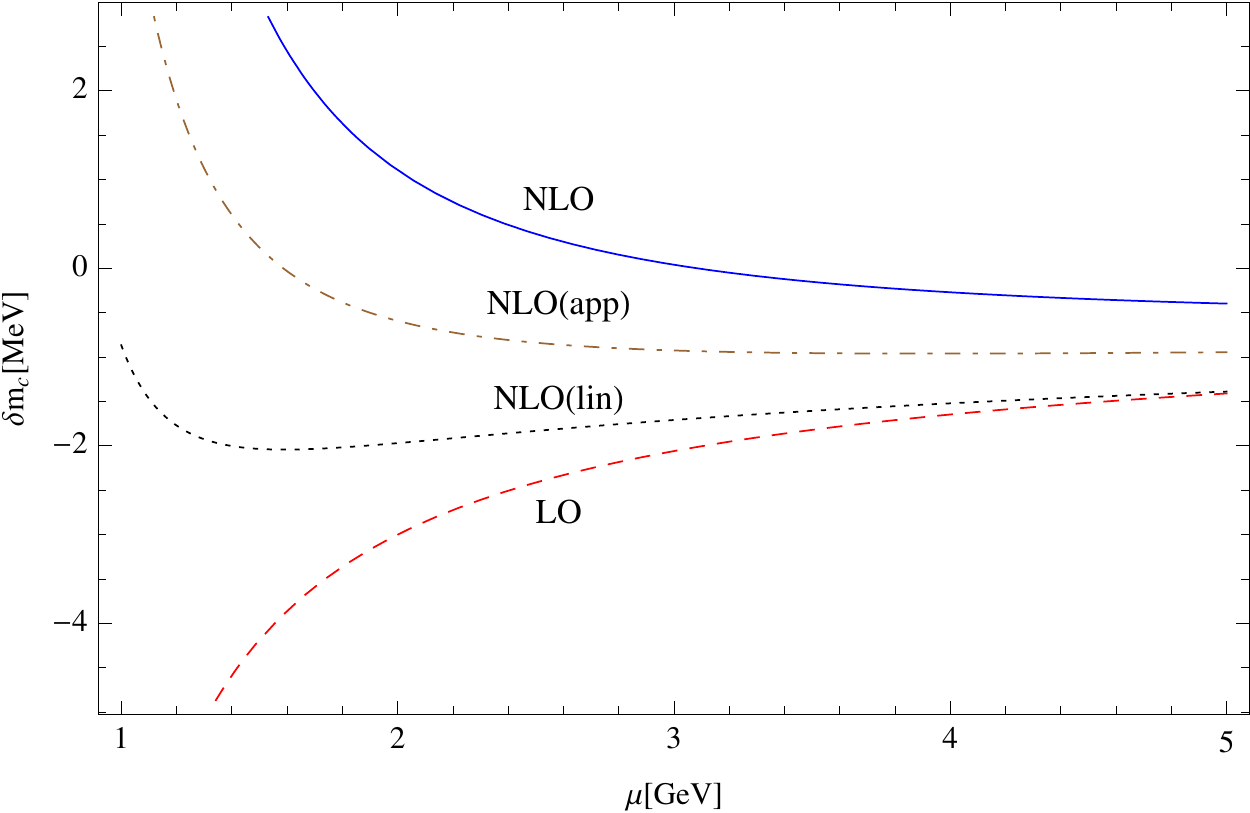}
%\centering\includegraphics[width=120mm]{delmcmuA.eps}
\vspace{-0.4cm}
\caption{\it Plot of the ${\cal O}(a_-^2)$ (dashed line) and ${\cal O}(a_-^3)$ (solid line) terms of $\delta m_c$ (cf. Eq.~(\ref{delmc})) 
as a function of the factorization scale. We also plot the ${\cal O}(a_-^3)$ term with $\delta m_{(c,+)}^{(2)}$ approximated 
to $\delta m_{(c,+)}^{(2,\rm lin)}$ (dotted line) and $\delta m_{(c,+)}^{(2,\rm app)}$ (dashed-dotted line).
}
\label{fig:deltamc}
 \end{figure}

\begin{figure}[htb] %\unitlength=1mm
%\centering\epsfig{file=mb(mu).eps,width=8.cm}
\centering\includegraphics[width=120mm]{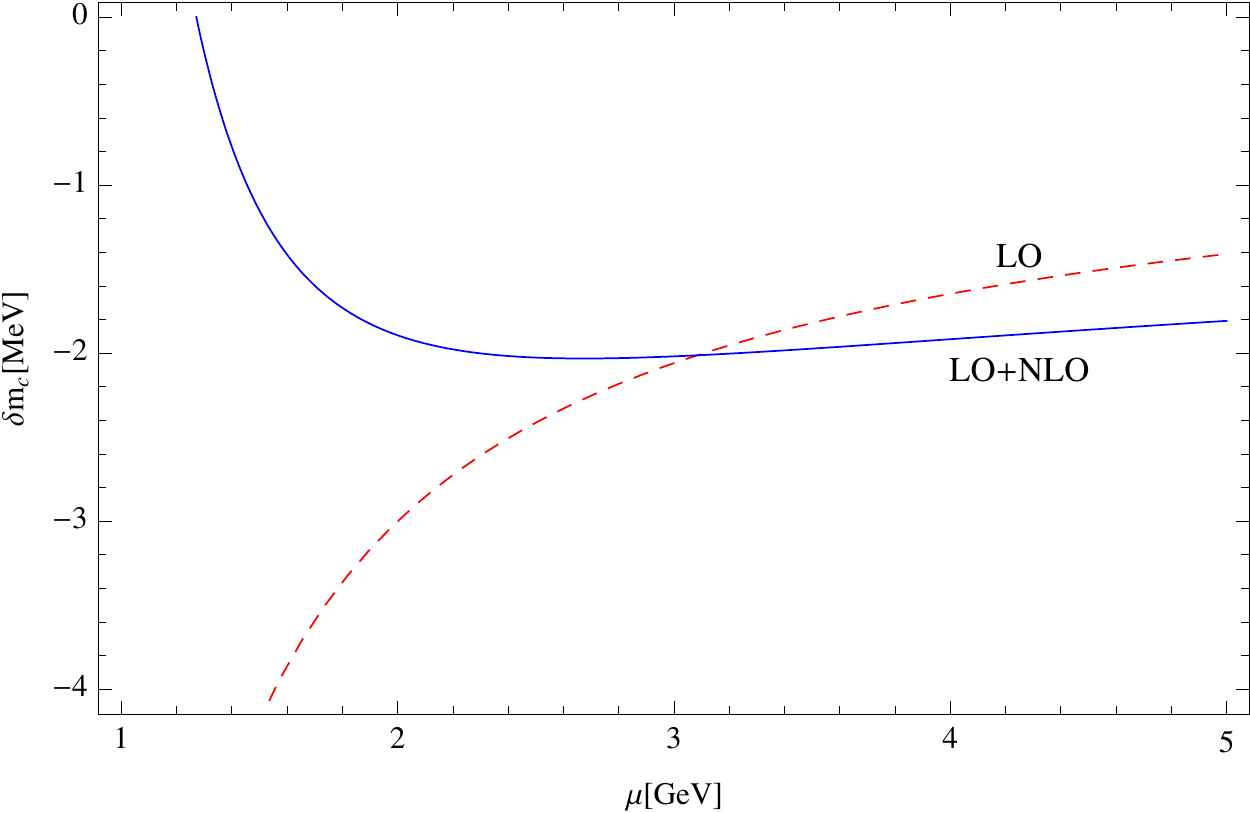}
%\centering\includegraphics[width=120mm]{delmcmuB.eps}
\vspace{-0.4cm}
\caption{\it Plot of $\delta m_c$ (cf. Eq.~(\ref{delmc})) truncated at ${\cal O}(a_-^2)$ (LO, dashed line) 
and ${\cal O}(a_-^3)$ (LO+NLO, solid line) as a function of the factorization scale.}
\label{fig:deltamcfinal}
 \end{figure}

If we put numbers we obtain $\left[\delta m_{(c,+)}^{(1)}+\delta m_{(c,\rm dec.)}^{(1)}
\right]a_{-}^{2}(\m_b)= -1.6$ MeV and $\left[\delta m_{(c,+)}^{(2)}+\delta m_{(c,\rm dec.)}^{(2)}
\right]a_{-}^{3}(\m_b)= -0.3$ MeV.  We observe that the series is now convergent, 
and the strong cancellation between $\delta m_{(c,+)}^{(i)}$ and $\delta m_{(c,\rm dec.)}^{(i)}$, as 
expected. This cancellation and convergence holds for different factorization scales, as we illustrate 
in Fig.~\ref{fig:deltamc} by comparing the absolute size of the LO (dashed line) and NLO (solid line) correction. 
This analysis makes clear that the magnitude of the ${\cal O}(a^2)$ charm effect is $\sim -2$ MeV (compared with $\sim \pm 10$ MeV 
for the individual terms $\delta m_{(c,+)}^{(1)}a_{-}^{2}(\m_b)$ and $\delta m_{(c,\rm dec.)}^{(1)}a_{-}^{2}(\m_b)$), and somewhat smaller than 
$\pm 1$ MeV for the ${\cal O}(a^3)$ charm effects (compared with $\sim \pm 20$ MeV 
for the individual terms $\delta m_{(c,+)}^{(2)}a_{-}^{3}(\m_b)$ and $\delta m_{(c,\rm dec.)}^{(2)}a_{-}^{3}(\m_b)$). 
After the cancellation, the ${\cal O}(a^3)$
charm-mass effect is clearly negligible compared with other uncertainties. Note also that, even though 
$\delta m_{(c,+)}^{(2,\rm lin)}$ reproduces quite well the magnitude of $\delta m_{(c,+)}^{(2)}$, it does not well enough to get an accurate 
value of the NLO correction after the cancellation. Therefore, the linear approximation could only be used to get the order of magnitude of the NLO 
effect (once the cancellation has been incorporated in the computation). We show this effect in Fig.~\ref{fig:deltamc} by comparing the exact 
NLO (solid line) correction with the linear approximation of the NLO correction (dotted line).  Note also that the precision required is such 
that $\delta m_{(c,+)}^{(2,\rm app)}$ is not accurate enough to reproduce the linear approximation of the NLO correction (compare the dotted and dashed-dotted lines in Fig.~\ref{fig:deltamc}). Finally, in Fig.~\ref{fig:deltamcfinal}, we give our final results for the charm-related contributions. Observe the smallness of the correction and the scale stability of the final result, producing a shift $\sim -2$ MeV. 

\subsection{Charm quark effects in the static potential}
\label{sec:charmVs}

In this subsection we directly work with $N_l$ massless active quarks (as motivated by the analysis of Ref.~\cite{Brambilla:2001qk}). The effects of the charm quark are included as an explicit correction to the potential.

The perturbation expansion of the QCD $q$-$\overline{q}$
static singlet potential is known with high accuracy. Its ${\cal O}(a^2)$ contribution was obtained in Ref.~\cite{Fischler:1977yf},
the ${\cal O}(a^3)$ in Refs.~\cite{Peter:1996ig,Schroder:1998vy}, the ${\cal O}(a^4)$ logarithmic term in 
Ref. \cite{Brambilla:1999qa}, the ${\cal O}(a^4)$ light-flavour finite piece in Ref. 
\cite{Smirnov:2008pn}, and the ${\cal O}(a^4)$ pure gluonic finite piece in Refs. \cite{Anzai:2009tm,Smirnov:2009fh}. In momentum space the potential reads
\bea
\lefteqn{
V(|\mathbf{k}|)=-\frac{16\pi^2}{3}\frac{1}{|\mathbf{k}|^2} a_-(\mu) 
{\bigg\{} 
1+ a_-(\mu)\left[\frac{1}{4}a_1+\beta_0 L \right]+a_-^2(\mu)\left[\frac{1}{4^2}a_2+\left(\frac{1}{2}a_1\beta_0+\beta_1 \right)L+\beta_0^2L^2 \right]
}
\nonumber\\
&&+a_-^3(\mu)\left[\frac{1}{4^3}a_3+b_3 {\rm ln}\left(\frac{\mu_f^2}{|\mathbf{k}|^2} \right)+\left(\frac{3}{16}a_2\beta_0+\frac{1}{2}a_1\beta_1+\beta_2 \right)L+\left(\frac{3}{4}a_1\beta_0^2+\frac{5}{2}\beta_0\beta_1 \right)L^2+\beta_0^2 L^3 \right]
\nonumber\\
&& +O(a_-^4)  {\bigg \}},
\label{Vk}
\eea
where $L={\rm ln}(\mu^2/|\mathbf{k}|^2)$ and $\mu$ is the renormalization scale. 
The coefficients $a_1$, $a_2$ and $a_3$ read
\bea
a_1&=& \frac{31}{3}-\frac{10}{9}N_l \ ,
\label{a1}
\\
a_2&=&\frac{100 }{81}N_l{}^2-\left(\frac{52 \zeta (3)}{3}+\frac{1229}{27}\right)N_l
+
9 \left(\frac{4343}{162}+\frac{1}{4} \left(16 \pi ^2-\pi ^4\right)+\frac{22\zeta (3)}{3}\right)
\ ,
\label{a2}
\\
a_3&=&a_3^{(0)}+a_3^{(1)} N_l+a_3^{(2)} N_l^2+a_3^{(3)} N_l^3\ ,
\label{a3}
\eea
where the coefficients in $a_3$ are
%\begin{subequations}
%\label{a3co}
\bea
\label{a3co}
a_3^{(0)}=
13432.6
%502.24 C_A^3-136.39\left(\frac{N_C\left(N_C^2+6\right)}{48}\right)
\ ,
%\label{a30}
%\\
a_3^{(1)}=
-3289.91
%-709.717 C_A^2T_F+\left(\frac{-71281}{162}+264 \zeta (3)+80\zeta (5)\right)C_A C_FT_F
%\nonumber
%\\
%&&
%+\left(\frac{286}{9}+296\frac{\zeta (3)}{3}-160\zeta
%(5)\right)C_F^2 T_F-56.83\left(\frac{18-6N_C^2+N_C^4}{96N_C^2}\right)
\ ,
%\label{a31}
%\\
a_3^{(2)}=
\frac{412 \zeta (3)}{9}+\frac{93631}{972}+\frac{16 \pi ^4}{45}
%C_FT_F^2 \left(\frac{14002}{81}-\frac{416\zeta (3)}{3}\right)+C_AT_F^2 \left(\frac{12541}{243}+\frac{64 \pi ^4}{135}+\frac{368 \zeta
%(3)}{3}\right)
\ ,
%\label{a32}
%\\
a_3^{(3)}=-\frac{1000}{729} \ .
%\label{a33}
\eea
%\end{subequations}

The terms involving powers of $L$ in Eq.~(\ref{Vk}) cancel the $\mu$ dependence of $\als(\mu)$ in $V$. Besides, at ${\cal O}(a^4)$
there is a factorization scale dependence that can not be absorbed in 
$\als$. We have singled out this contribution, it is proportional to 
\be
b_3=27\pi^2/24 \ ,
\label{b3}
\ee
and depends on the infrared cutoff $\mu_f$, which cuts out ultrasoft ($us$) degrees of freedom ($|k^2|^{1/2}\sim m_q\alpha_s^2\sim E_{us}$) from the potential, which is characterized by the soft scale ($|k^2|^{1/2}\sim m_q\alpha_s\sim E_{s}$): $E_{us}<\mu_f<E_{s}$. The existence of the infrared divergent terms at $\sim a^4$ in the static Wilson loop was first pointed out in Ref.~\cite{Appelquist:1977es}.  

The three-dimensional Fourier transformation of Eq.~(\ref{Vk}) gives the perturbation 
expansion of the static potential in position space
\bea
V(r)&=&-\frac{4\pi}{3}\frac{1}{r} a_-(\mu) {\bigg \{}
1+ a_-(\mu)\left[\frac{1}{4}a_1+2 \beta_0 l \right]
\nonumber
\\
&&+a_-^2(\mu)\left[\frac{1}{4^2}a_2+\left(a_1\beta_0+2\beta_1 \right)l+\beta_0^2(4 l^2+\pi^2/3) \right]
\nonumber
\\
&&+a_-^3(\mu) {\bigg [} \frac{1}{4^3}a_3+2 b_3 {\rm ln}\left(\mu_f r {\rm e}^{\gamma_E} \right)+\left(\frac{3}{16}a_2\beta_0+\frac{1}{2}a_1\beta_1+\beta_2 \right)2l
\nonumber
\\
&&+\left(\frac{3}{4}a_1\beta_0^2+\frac{5}{2}\beta_0\beta_1 \right)(4 l^2+\pi^2/3)+\beta_0^3 (8 l^3+2\pi^2 l+16\zeta(3)) {\bigg ]} +O(a_-^4) {\bigg \}} \ ,
\label{Vr}
\eea   
where the notation $l={\rm ln}(\mu r \exp(\gamma_E))$ is used, 
with $\gamma_E$ being the Euler constant ($\gamma_E=0.5772...$).

The leading charm quark correction to the potential is the following
\begin{equation}
\delta V_c(r) = -\frac{4\pi}{3}
\frac{a_-(\mu)}{r}\left(\frac{a_-(\mu)}{3}
\right) \int_{1}^{\infty}dx\,
\frac{\sqrt{x^{2}-1}}{x^{2}}\left(1+\frac{1}{2x^{2}}
\right) e^{-2m_{c}rx}\,.
\label{eq:vpmc2}
\end{equation}
Its effect will be quite tiny. Therefore, we have only incorporated 
Eq.~(\ref{eq:vpmc2}) in our final evaluations and have not considered any other subdominant effects in the charm mass. 

\section{Leading renormalon of the pole mass and the singlet static potential}
\label{TB}

The determination of the normalization constant of the leading infrared renormalon of the pole mass (and the singlet static potential) 
is an essential ingredient for the RS scheme defined in Ref.~\cite{Pineda:2001zq}. Therefore, in this section we want to improve over previous determinations of this quantity.

It is clear from the discussion of Sec. \ref{sec:charm} that the charm quark decouples at large orders in perturbation theory (in practice this happens at rather low orders). Therefore, we work in the theory with $N_l$ ($N_l=3$ for the bottom case) active (masless flavours), and all the coefficients in this section should be understood with $N_f=N_l$.

The leading asymptotic behaviour of the perturbation series of $m_b$ (see Eq. (\ref{SmNl}))
is determined by 
the leading infrared renormalon ambiguity of $m_b$. 
This ambiguity $\delta m_b$ is renormalization scale and
scheme independent and is a QCD scale with the dimension 
of energy; therefore, it must be proportional to the QCD scale 
$\Lambda_{\rm QCD}$: $\delta m_b = const \times \Lambda_{\rm QCD}$ 
\cite{Beneke:1994rs}. This scale, written in terms of
$a(\mu)$ and of the renormalization scale $\mu$,
has the form
\begin{equation}
\Lambda = const \times \mu \exp \left( - \frac{1}{2 \beta_0 a(\mu)}
\right) a^{- \nu}(\mu) c_1^{- \nu} \left[
1 + \sum_{k=1}^{\infty} \ (2 \beta_0)^k \nu (\nu-1) \cdots
(\nu-k+1){\widetilde c}_k a^k(\mu) \right] \ ,
\label{tL2}
\end{equation}
where
\begin{subequations}
\label{tr}
\begin{eqnarray}
\nu & = & \frac{c_1}{ 2 \beta_0} = \frac{\beta_1}{2 \beta_0^2} \ ,
\label{nu}
\\
{\widetilde c}_1 & = & \frac{ ( c_1^2 - c_2) }{(2 \beta_0)^2 \nu} \ ,
\qquad
{\widetilde c}_2 = \frac{1}{2 (2 \beta_0)^4 \nu (\nu\!-\!1)} \left[
( c_1^2 - c_2 )^2 - 2 \beta_0 ( c_1^3 - 2 c_1 c_2 + c_3 ) \right] \ .
\label{tr1r2}
\\
{\widetilde c}_3 & = & \frac{1}{6 (2 \beta_0)^6 \nu (\nu\!-\!1)(\nu\!-\!2)} 
\left[
( c_1^2 - c_2 )^3 - 6 \beta_0 (c_1^2 - c_2 )(c_1^3 - 2 c_1 c_2 + c_3 )
\right.
\\
\nn
&&
\left.
+ 8 \beta_0^2 ( c_1^4 - 3 c_1^2 c_2 + c_2^2 + 2 c_1 c_3 - c_4 )
\right] .
\label{tr3}
\end{eqnarray}
\end{subequations}

The renormalon ambiguities give us information on the asymptotic 
behaviour of the perturbation expansion. 
This information is easily encoded in the Borel transform of $S$, 
\begin{equation}
B_{S}(u; \mu) \equiv \frac{4}{3} \left[ 1 + \frac{r_1(\mu)}{1! \; \beta_0} u +
\frac{r_2(\mu)}{2! \; \beta_0^2} u^2  +
\frac{r_3(\mu)}{3! \; \beta_0^3} u^3 + {\cal O}(u^4) \right] \ .
\label{BSm1}
\end{equation}
This function has renormalon singularities at 
$u = 1/2, 3/2, 2, \ldots, -1, -2, \ldots$ 
\cite{Bigi:1994em,Beneke:1994sw,Beneke:1999ui}, and likely also at $u=+1$~\cite{Neubert:1996zy}.
Except for the normalization, the leading infrared renormalon ambiguity of $m_b$, $\delta m_b = const \times \Lambda_{\rm QCD}$, completely determines \cite{Beneke:1994rs} the behaviour of the nearest singularity to the origin (at $u=1/2$) of $B_S$, since 
$\Lambda = 
\kappa {\rm Im} S_{\rm BI}(z=2 \beta_0 a(\mu)- i \epsilon)$,
where $\kappa$ is a $\mu$-independent constant and
`BI' denotes the Borel-integrated expression for $S$.\footnote{
For an explicit expression for ${\rm Im} S_{\rm BI}(z=2 \beta_0 a(\mu)- i \epsilon)$,
see, for example, Ref.~\cite{Cvetic:2002qf}.} We then have
$B_S(u; \mu)$
\begin{eqnarray}
B_S(u; \mu) & = &  N_m \pi  \frac{\mu}{ {\overline m}_b }
 \frac{1}{ ( 1 - 2 u)^{1 + \nu} } \left[ 1 +
\sum_{k=1}^{\infty} {\widetilde c}_k ( 1 - 2 u)^k \right]
+ B_{S}^{\rm (an.)}(u; \mu) \ ,
\label{BSrenan}
\end{eqnarray}
where $N_m$ is the residue (normalization constant) parameter of the
renormalon. 
$\tilde c_1$ was first computed in Ref.~\cite{Beneke:1994rs}, and 
$\tilde c_2$ in Refs.~\cite{Beneke:1999ui,Pineda:2001zq}. 
We also give a value for ${\widetilde c}_3$ 
using  the estimate for the ${\overline {\rm MS}}$ scheme coefficient $c_4 = \beta_4/\beta_0$
obtained in Ref.~\cite{Ellis:1997sb}
by Pad\'e-related methods  
\be
\beta_4= \frac{1}{4^5} (A_4 + B_4 N_f + C_4 N_f^2 + D_4 N_f^3 + E_4 N_f^4)
\,,
\label{beta4}
\ee 
where $A_4= 7.59 \times 10^5$, $B_4= -2.19 \times 10^5$, $C_4= 2.05 \times 10^4$,
$D_4= -49.8$, and $E_4=-1.84$. This results in $c_4=123.7$ for $N_f=3$,
$c_4=97.2$ for $N_f=4$, and $c_4=86.2$ for $N_f=5$. 

$B_{S}^{\rm (an.)}(u; \mu)$ is analytic on the disk $|u| < 1$. 
Therefore, even in the vicinity of $u \sim 1/2$
 it can be expanded in powers of $u$
\be
B_{S}^{\rm (an.)}(u; \mu) =  h_0(\mu) + 
\sum_{N \geq 1} \frac{h_N(\mu)}{N! \; \beta_0^N} u^N \ .
\label{BSanexp}
\ee
The coefficients $h_N(\mu)$
of the analytic part (\ref{BSanexp}) are exponentially suppressed in $N$, $  \sim e^{-N}$, 
in comparison with the large coefficients $r_N(\mu)$. Their relation is
obtained by equating the expansion of Eq.~(\ref{BSrenan}) in powers of $u$
with the expansion (\ref{BSm1}). This gives
\be
\label{hms}
 \frac{4}{3} r_N(\mu)  =    \pi N_m \frac{\mu}{{\overline m}_b}
(2 \beta_0)^N \sum_{s \geq 0} {\widetilde c}_s 
\frac{ \Gamma ( \nu + N + 1 - s) }{\Gamma(\nu + 1 - s) } +h_N(\mu)
\ ,
\ee
where we recall that $r_0 = {\widetilde c}_0 = 1$. 
The sum in Eq.~(\ref{hms}) introduces ${\cal O}(1/N)$ corrections
to the leading asymptotic behaviour. 
The numbers ${\widetilde c}_s$
entering  the sum in Eq.~(\ref{hms}), are given by Eqs.~(\ref{tr})
and are known for $s \leq 3$.\footnote{
For $N_f=3$:
${\widetilde c}_1= -0.1638$, ${\widetilde c}_2= 0.2372$, ${\widetilde c}_3= -0.1205$;
$\nu= 0.3951$. 
For $N_f=4$: ${\widetilde c}_1= -0.1054$, ${\widetilde c}_2= 0.2736$, 
${\widetilde c}_3= -0.1610$; $\nu= 0.3696$.
For $N_f=5$:
${\widetilde c}_1= 0.0238$, ${\widetilde c}_2= 0.3265$, ${\widetilde c}_3= -0.2681$;
$\nu= 0.3289$.}
Therefore, by default, we truncate the sum in (\ref{hms}) at $s=3$. This introduces an error of
order ${\cal O}(1/N^4)$ for the asymptotic behaviour (we will typically take the difference between truncating the
sum at $s=2$ or $s=3$ to check the quality of the approximation). We also set $h_N=0$ since 
they yield (in comparison) exponentially suppressed terms. Overall, we approximate the asymptotic behaviour 
of $r_N$ by the following equality: 
\bea
\label{rNasym}
&&
 \frac{4}{3} r^{asym}_N(\mu)  \simeq   \pi N_m \frac{\mu}{{\overline m}_b}
(2 \beta_0)^N 
\frac{ \Gamma ( \nu + N + 1) }{\Gamma(\nu + 1) } 
\\
&&
\qquad
\nn
\times
\left(
1+\frac{\nu}{N+\nu}{\widetilde c}_1+\frac{\nu(\nu-1)}{(N+\nu)(N+\nu-1)}{\widetilde c}_2
+\frac{\nu(\nu-1)(\nu-2)}{(N+\nu)(N+\nu-1)(N+\nu-2)}{\widetilde c}_3+{\cal O}(\frac{1}{N^4})
\right)
\ .
\eea

\medskip

We can have a similar discussion for the static potential. 
The Borel transformation of the dimensionless potential $(-\frac{3}{4\pi})rV(r)$ in (\ref{Vr}) is given by:
\be
B_V(u,\mu)=1+\frac{v_1}{1!\beta_0}u+\frac{v_2}{2!\beta_0^2}u^2+\frac{v_3}{3!\beta_0^3}u^3+\ldots,
\label{BF}
\ee 
where $v_j$ is the coefficient at the power $a^j(\mu)$ 
in the expansion (\ref{Vr}) ($v_1=a_1/4+2\beta_0 l, etc.$). This function has the renormalons located at $u=1/2,3/2,5/2$, etc., Ref.~\cite{Aglietti:1995tg}. It can be written as
\be
B_V(u,\mu r)=-\frac{3}{4}N_V \mu r \frac{1}{(1-2u)^{1+\nu}}\left[1+\sum_{k=1}^{\infty} {\widetilde c}_k ( 1 - 2 u)^k \right]+({\rm analytic \ term}) \ .
\label{BF2}
\ee 
The expression in the brackets is $\mu$-independent, and the last term is analytic for $|u|<3/2$.

The asymptotic behaviour of $v_N$ is equal to the behaviour of $r_N$ except for the normalization:
\be
- \frac{4}{3} v_N(\mu)  =    N_V \mu r
(2 \beta_0)^N \sum_{s \geq 0} {\widetilde c}_s 
\frac{ \Gamma ( \nu + N + 1 - s) }{ \Gamma(\nu + 1 - s) } +d_N(\mu)\ .
\label{vN}
\ee
The coefficients $d_N$ are 
analogous to the coefficients $h_N$ for the pole mass. We will set them equal
to zero 
for the same reason, as they yield exponentially suppressed terms in comparison. We also truncate the sum to the
first known terms ($s \leq 3$). Therefore, we approximate the asymptotic behaviour 
of $v_N$ by the following equality: 
\bea
\label{vNasym}
&&
 -\frac{4}{3} v^{asym}_N(\mu)  \simeq  N_V \mu r
(2 \beta_0)^N 
\frac{ \Gamma ( \nu + N + 1) }{\Gamma(\nu + 1) } 
\\
&&
\qquad
\nn
\times
\left(
1+\frac{\nu}{N+\nu}{\widetilde c}_1+\frac{\nu(\nu-1)}{(N+\nu)(N+\nu-1)}{\widetilde c}_2
+\frac{\nu(\nu-1)(\nu-2)}{(N+\nu)(N+\nu-1)(N+\nu-2)}{\widetilde c}_3+{\cal O}(\frac{1}{N^4})
\right)
\ .
\eea

\subsection{Determination of $N_V$}
\label{sub:p}

In this subsection we compute $N_V$ using two methods that we name A) and B). 

\begin{figure}[htb] %\unitlength=1mm
%\centering\epsfig{file=mb(mu).eps,width=8.cm}
\centering
\includegraphics[width=140mm]{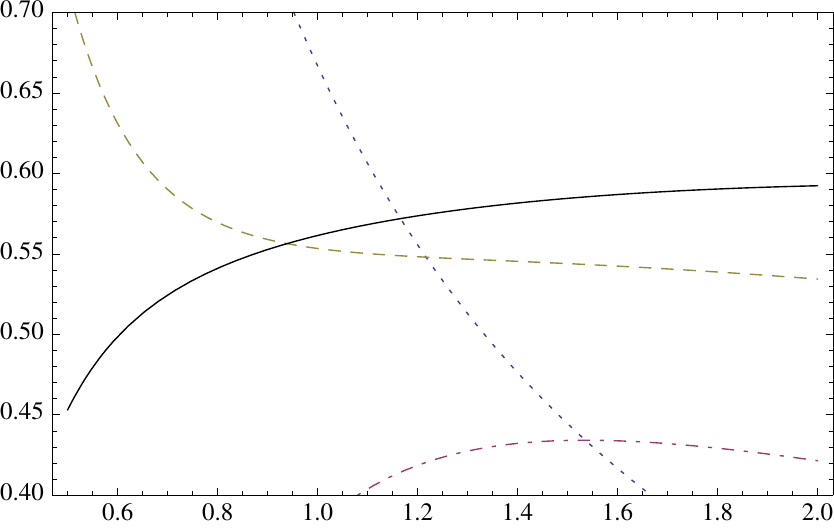}
\put(-150,70){{\Large LO}}
\put(1,20){{\Large NLO}}
\put(1,108){{\Large NNLO}}
\put(1,158){{\Large N{}$^3$LO}}
\put(10,5){{\Large $x$}}
\put(-460,240){{\Large $-N_V/2$}}
\vspace{-0.2cm}
\caption{\it Method A): $-N_V/2$ obtained using Eq.~(\ref{RVNV}) for $N_l=3$, as a function of $x \equiv \mu r$, truncated at $N=0,1,2,3$, which we name as
LO (dotted), NLO (dashed-dotted), NNLO (dashed) and NNNLO (solid) respectively.}
\label{fig:NVmethodA_nf3}
\end{figure}

The method A) uses the idea of Refs.~\cite{Lee:1996yk,Lee:1999ws}
of, instead of working with Eq.~(\ref{BF2}), using an associated function that kills the leading singularity in the Borel plane. 
This idea was first applied to the static potential (and the pole mass) in Ref.~\cite{Pineda:2001zq} 
and also used in \cite{Pineda:2002se,Brambilla:2010pp,Cvetic:2003wk}. 
One uses 
\be
R_V(u;\mu r)\equiv  -\frac{4}{3}
\frac{1}{\mu r}(1-2u)^{1+\nu}B_V(u,\mu r)=\sum_{k=0}^{\infty} R^{(k)}_V u^k
\,,
\label{RVform}
\ee 
which is defined such that the leading singularity at $u=1/2$ of $B_V$ is eliminated. Its evaluation at $u=1/2$ 
gives
\be
N_V=R_V(u=1/2;\mu r) \simeq \sum_{k=0}^{N} R^{(k)}_V(\mu r) \left(\frac{1}{2}\right)^k
\,.
\label{RVNV}
\ee
In this paper we carefully study the case when the sum is truncated at $N=3$. Truncating the infinity sum to its first orders produces some remaining scale dependence. In particular, from $N=3$ on a dependence on the ultrasoft factorization scale 
$\mu_f$ appears, which we set equal to $\mu$. We plot the scale dependence of $-N_V/2$ (the relevant quantity to be compared with $N_m$) in Fig. \ref{fig:NVmethodA_nf3} for $N=0,1,2,3$ for $N_l=3$. 
We observe a nicely convergent pattern, specially for the difference between the $N=2$ and $N=3$ computation. Note also that 
we expect the results to be better for $\mu r \sim 1$, since $\ln (\mu r)$ terms are not large.

\begin{figure}[htb] %\unitlength=1mm
%\centering\epsfig{file=mb(mu).eps,width=8.cm}
\centering
\includegraphics[width=140mm]{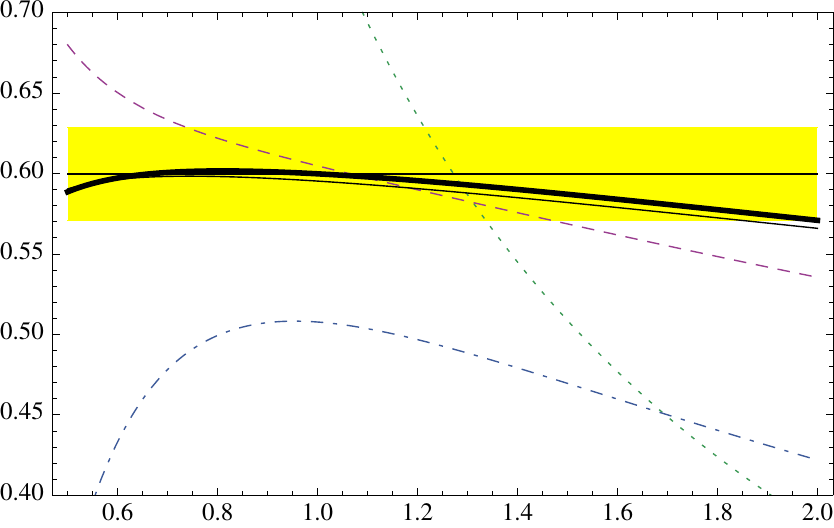}
\put(-100,80){{\Large LO}}
\put(1,25){{\Large NLO}}
\put(1,110){{\Large NNLO}}
\put(1,145){{\Large N{}$^3$LO}}
\put(10,-5){{\Large $x$}}
\put(-455,240){{\Large $-N_V/2$}}
\vspace{-0.cm}
\caption{\it Method B): $-N_V/2$ for $N_l=0$, as a function of $x \equiv \mu r$, 
obtained from $-(N_V/2)v_N/v_N^{asym}$. $v_N$ is taken from Eq. (\ref{Vr}) and
$v_N^{asym}$ from Eq. (\ref{vNasym}) truncated at ${\cal O}(1/N^3)$. 
We name the 
different lines as LO (dotted), NLO (dashed-dotted), NNLO (dashed) and NNNLO (thick solid) 
for $N=0,1,2,3$, respectively. 
The horizontal line and the central band is our final estimate for $-N_V/2$ and its error in Eq. (\ref{eq:NV}). 
We also plot the NNNLO curve without the US term (thin solid line).}
\label{fig:Nmnf0_V}
\end{figure}
 
\begin{figure}[htb] %\unitlength=1mm
%\centering\epsfig{file=mb(mu).eps,width=8.cm}
\centering
\includegraphics[width=140mm]{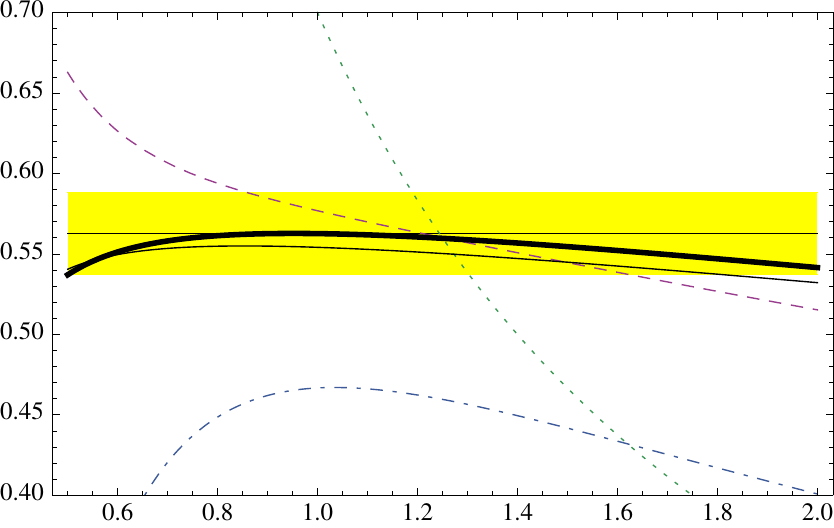}
\put(-130,70){{\Large LO}}
\put(-170,30){{\Large NLO}}
\put(1,94){{\Large NNLO}}
\put(1,119){{\Large N{}$^3$LO}}
\put(10,5){{\Large $x$}}
\put(-455,240){{\Large $-N_V/2$}}
\vspace{-0.2cm}
\caption{\it Same as in Fig.~\ref{fig:Nmnf0_V}, but now for $N_l=3$.
}
\label{fig:Nmnf3_V}
\end{figure}
 
The method B) determines $N_V$ by dividing the exactly known coefficients $v_N$ 
directly obtained from Eq. (\ref{Vr}) 
by the large $N$ renormalon-based expectations (cf. Eq. (\ref{vNasym})), which we truncate at ${\cal O}(1/N^3)$, including the $\sim 1/N^3$ terms.
If we are reaching the asymptotic regime we should converge to a constant and get a mild scale dependence. We plot the results in Figs. \ref{fig:Nmnf0_V} and \ref{fig:Nmnf3_V} for $N_l=0$ and $N_l=3$ respectively. We observe a very nice convergence, with a milder scale dependence as we go to higher orders. We observe a sizable effect of the $1/N^3$  
truncation for $N=0,1$ (actually the $N=1$ result is closer to the asymptotic result if we truncate at 
$1/N^2$ than at $1/N^3$ order) 
but negligible for $N=2$ and $N=3$.

If we compare method A) and B), we observe that method B) yields a more 
convergent series and a milder scale dependence for $N=3$ than method A). Therefore, we fix the central value of $N_V$ (for all values of $N_f$) from the result obtained from method B) for $N=3$ and $x=\mu r=\mu_f r=1$. We also use method B) to fix the error of this determination: we compute the difference of our central value with the evaluations with $x\in [1/2,2]$ and the difference between the NNLO and NNNLO result at $x=1$ and take the maximum of the two as our error estimate. To this error we add in quadrature the difference of the NNNLO result obtained truncating the asymptotic expression at ${\cal O}(1/N^3)$ or at 
${\cal O}(1/N^2)$. We find this effect to be way subleading in 
comparison with the scale variation. Finally, for $N_l=0,3$ we obtain
\be
\label{eq:NV}
-\frac{N_V}{2}\Bigg|_{N_l=0}= 0.600(29) \,, \qquad  -\frac{N_V}{2}\Bigg|_{N_l=3}= 0.563(26)
\,.
\ee 
As expected, determinations with $x \equiv \mu r \sim 1$ yield the best results, since this minimizes possible large $\ln (\mu r)$ terms. 

The factorization scale $\mu_f$ is not related with the leading infrared renormalon. Eliminating this contribution altogether allows us to measure the quality of our error estimate. We plot the determination of $N_V$ if we completely eliminate the ultrasoft term in Eq. (\ref{Vr}) in Figs. \ref{fig:Nmnf0_V} and \ref{fig:Nmnf3_V}. We observe that the associated shift is much smaller than the scale variation of the NNNLO curve, specially for the $N_l=0$ case.

The fact that method B) yields a more convergent (and stable) series was also clearly observed in Ref.~\cite{Bali:2013pla}, where 
$N_m$ was determined from the perturbative computation of the self-energy of 
a static source to ${\cal O}(\als^{20})$ 
(compare Fig. 12 with Fig. 14 in this reference).
 
\subsection{Determination of $N_m$ from the pole mass}
\label{sub:hm}

\begin{figure}[htb] %\unitlength=1mm
%\centering\epsfig{file=mb(mu).eps,width=8.cm}
\centering
\includegraphics[width=140mm]{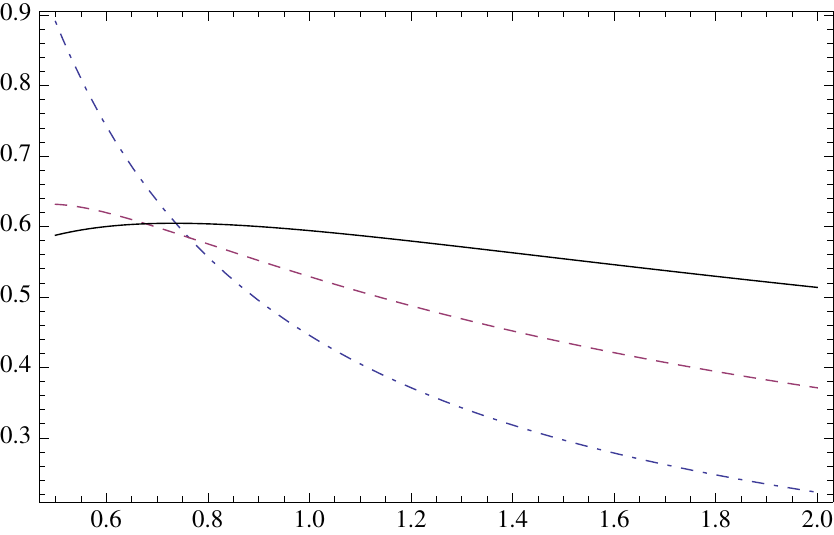}
\put(1,10){{\Large LO}}
\put(1,62){{\Large NLO}}
\put(1,110){{\Large NNLO}}
\put(10,-5){{\Large $x$}}
\put(-430,240){{\Large $N_m$}}
\vspace{-0.cm}
\caption{\it 
Method B): $N_m$ for $N_l=3$, as a function of $x \equiv \mu/\m_b$, 
obtained from $r_N/r_N^{asym}$. $r_N$ is taken from Eq. (\ref{rs}) and
$r_N^{asym}$ from Eq. (\ref{rNasym}) truncated at ${\cal O}(1/N^3)$. 
We name the 
different lines as NLO (dashed-dotted), NLO (dashed) and NNLO (solid) 
for $N=0,1,2$, respectively. }
\label{fig:Nmnf3_mOS}
\end{figure}

In this subsection we obtain $N_m$ from the perturbation 
expansion of the pole mass using the methods A) and 
B) described in the previous section.  

From method A) one obtains $N_m$ from
\begin{equation}
N_m = \frac{{\overline m}_b}{\mu} \frac{1}{\pi} R_S(u=1/2; \mu) 
\simeq \sum_{n=0}^{N} R^{(n)}_S(\mu r) \left(\frac{1}{2}\right)^n
\ ,
\label{Nmform}
\end{equation}
where
\be
R_S(u; \mu) \equiv (1 - 2 u)^{1 + \nu} B_S(u; \mu)
\label{RS}
\ee
Truncating the sum to $N=2$, we recover the results of 
Refs.~\cite{Pineda:2001zq,Lee:2003hh,Cvetic:2003wk}. 
Unlike in the previous section, we can not go to one order higher since the  ${\cal O}(a^4)$
term of the pole-$\MS$ mass relation is not known. Therefore, we do not dwell further on this method.

With method B) we determine $N_m$ by dividing the exactly known coefficients $r_N$ (cf. Eq. (\ref{rs}))
by the large $N$ renormalon-based expectations, $r^{asym}_N$ (cf. Eq. (\ref{rNasym})),
which we truncate at ${\cal O}(1/N^3)$, including the $\sim 1/N^3$ terms.
We show the result in Fig.~\ref{fig:Nmnf3_mOS}. Again the maximum 
possible accuracy is at the highest $N$, this time
$N=2$. Overall, we know the perturbative relation between the pole and 
$\MS$ to one order less than in the case of the static potential. Therefore, the predictions in this
section are generically less precise.
 
\subsection{Final determination of $N_m$}
\label{Sec:Nm}

In the situation where $1/r\gg \Lambda_{QCD}$, one can do the matching 
between NRQCD and pNRQCD in perturbation theory, and $2m_q+V(r)$ can be understood as an observable up to 
$O(r^2\Lambda_{QCD}^3,\Lambda_{QCD}^2/m)$ renormalon (and/or nonperturbative) contributions (see 
the discussion in Ref.~\cite{Pineda:2001zq}). This implies that the leading infrared renormalon of the singlet static potencial 
must cancel with the leading renormalon of twice the pole mass, so that the following relation between $N_m$ and $N_V$ holds:
\be
2 N_m + N_V = 0 \ .
\label{resrel}
\ee
Therefore, we have several alternative determinations of $N_m$ (or $N_V$) from the analysis of the previous subsections. We now study the quality of them and choose the optimal\footnote{One may think of other observables, the perturbation expansions of which are 
dominated by the pole mass renormalon. One of those is the anomalous magnetic moment of the heavy quark.
In Ref.~\cite{Grozin:1997ih} it was shown (see also Ref.~\cite{Pineda:2013lta}) that, $(1+\kappa)/m$ is renormalon free. More precisely, the leading renormalon of $1+\kappa$ 
cancels with the leading renormalon of $m$ and one can write $\frac{1+\kappa}{m}=\frac{1}{\bar m}\frac{\bar m(1+\kappa)}{m}\equiv
\frac{1}{\bar m}(1+C_{\kappa})$, where $C_{\kappa}=\sum_{n=0}C_{\kappa}^{(n)}\als^{n+1}$ is free of the $u=1/2$ renormalon. Therefore, we are in the situation where  
$\bar m(1+\kappa)=m(1+C_{\kappa})$ has the pole mass (renormalon) but modulated by a nontrivial Wilson coefficient. This changes the 
$1/N$ corrections of the asymptotic expression. We have studied this quantity and obtained a number for the normalization of the renormalon compatible with ours though less precise.}. 

In Sec. \ref{sub:hm} we have determined $N_m$ using what we have named method A) and B). 
The results from method A) are nothing but those obtained in Ref.~\cite{Pineda:2001zq}, where the estimated 
uncertainty was of around 10\%. The results from method B) are new, and summarized in Fig. \ref{fig:Nmnf3_mOS}. Both methods use the perturbation expansion of the pole mass 
 to  ${\cal O}(a^3)$, 
which is one order less than for the static potential. Actually, these fits typically yield a stronger scale dependence than for the case of the static potential. Therefore, we will not dwell further with these determinations.

\begin{table}
\caption{\label{tab:Nm}\it Final predictions for $N_m$ and $N_V$ for different values of $N_l$.}
\begin{tabular}{|c|c|c|c|c|c|c|c|} \hline
$N_f$ &  0 & 1 & 2 & 3 & 4 & 5&6\\ \hline
$N_m=-N_V/2$ & 0.600(29)& 0.588(27) & 0.576(24) & 0.563(26) & 0.547(33) & 0.527(51)& 0.500(152) \\ \hline
\end{tabular}
\end{table}

\begin{figure}[htb] %\unitlength=1mm
%\centering\epsfig{file=mb(mu).eps,width=8.cm}
\centering\includegraphics[width=160mm]{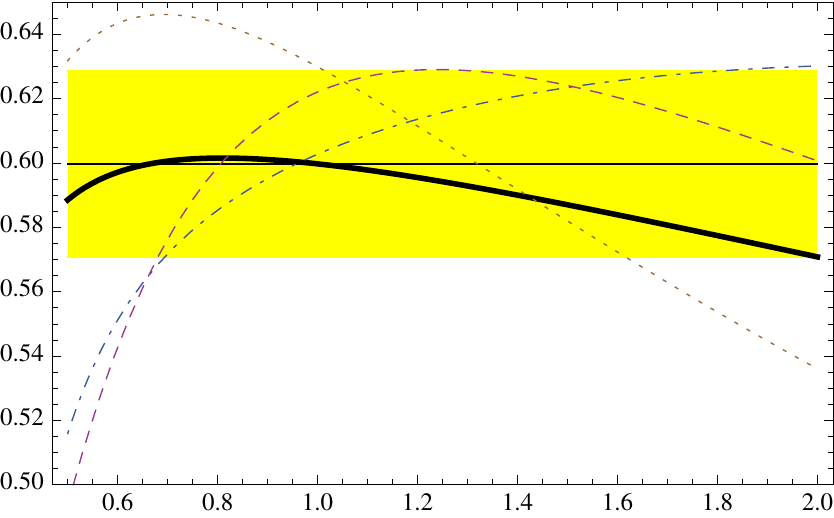}
%\centering\includegraphics[width=160mm]{Nmnf0_methods.eps}
\put(10,5){{\Large $x$}}
\put(-480,265){{\Large $N_m$}}
\vspace{-0.2cm}
\caption{\it $N_m$ obtained applying methods A) (dashed-dotted line)  and B) (solid thick curve)
to the the static potential at NNNLO  for $N_l=0$ 
as a function of $x \equiv \mu r$. For comparison we also include the NNLO evaluation from 
method A) and B) applied to the pole mass (dashed and dotted line respectively). 
The horizontal central line and bands correspond to our final 
central value and error.}
\label{fig:Nmnf0_methods}
\end{figure}

\begin{figure}[htb] %\unitlength=1mm
%\centering\epsfig{file=mb(mu).eps,width=8.cm}
\centering\includegraphics[width=160mm]{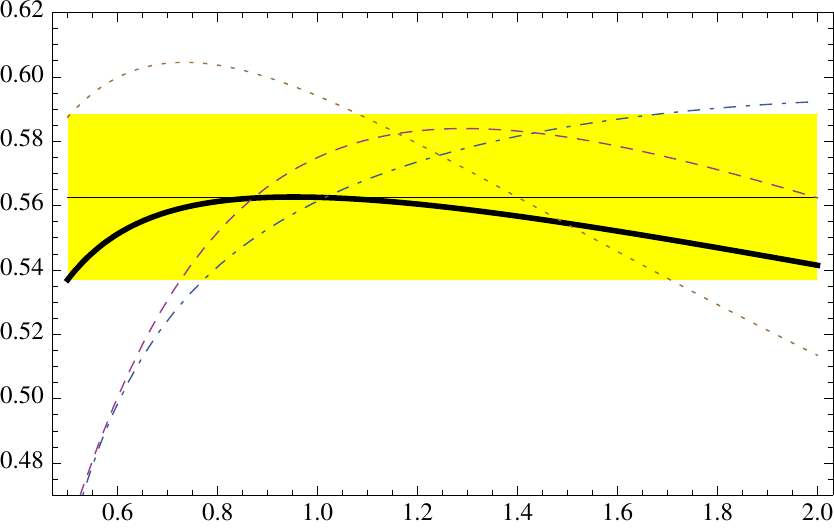}
%\centering\includegraphics[width=160mm]{Nmnf3_methods.eps}
\put(10,5){{\Large $x$}}
\put(-480,265){{\Large $N_m$}}
\vspace{-0.2cm}
\caption{\it Same as in Fig.~\ref{fig:Nmnf0_methods}, but for $N_{l}=3$. 
}
\label{fig:Nmnf3_methods}
\end{figure}

In Sec. \ref{sub:p}, we have obtained two new determinations of $N_V$ using the perturbation 
expansion of the static potential to  ${\cal O}(a^4)$, 
one order more than for the case of the pole mass. Therefore, 
we expect our new determinations to yield more accurate results. We have found this is specially so for method B of Sec. \ref{sub:p}. Since $N_V$ and $N_m$ are related by Eq. (\ref{resrel}), this gives a determination of $N_m$, which we take as the most precise and display in 
Table \ref{tab:Nm} as our final numbers. 
To illustrate this, in Figs. \ref{fig:Nmnf0_methods} and \ref{fig:Nmnf3_methods}, we compare the NNNLO evaluations using method B) (of the static 
potential) with method A) (of the static potential). We also include the NNLO evaluations using methods A) and B) (of the pole mass).
Around $x \sim 1$ all of them agree within one standard deviation. 
This signals 
that alternative errors estimates would give similar numbers.
We see how the NNNLO evaluation using method B yields the more stable result under scale variations. For the $N_{l}=0$ 
case we can 
compare with the value obtained in Ref.~\cite{Bali:2013qla}. We agree within one standard deviation. This is quite rewarding as these numbers have been obtained with completely different methods.

\begin{figure}[htb] %\unitlength=1mm
%\centering\epsfig{file=mb(mu).eps,width=8.cm}
\centering
\includegraphics[width=140mm]{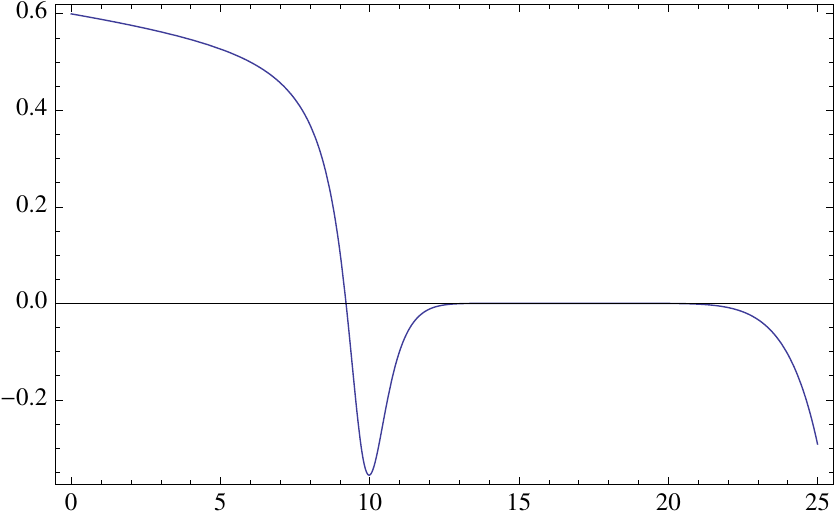}
\put(1,10){{\Large $N_l$}}
\put(-420,240){{\Large $N_m$}}
\vspace{-0.4cm}
\caption{\it $N_m(x=1)$ obtained using method B) 
from the static potential (NNNLO) as a function of $N_l$.}
\label{fig:Nmnf}
\end{figure}

It is also interesting to study the $N_l$ dependence of $N_m$. The $u=1/2$ infrared renormalon should disappear when $N_l \rightarrow \infty$, as 
the theory is not asymptotically free anymore. We plot $N_m$ as a function of $N_l$ (we fix $x=1$) using our preferred method (method B) from the 
static potential in Fig. \ref{fig:Nmnf}. We observe how $N_m$ tends to zero in the range of  $N_l \in (12,23)$, a range of values for which one could expect a conformal window. This shows the disappearance of $u=1/2$ infrared renormalon for $N_l >12$.  In the range $N_l \in (25,40)$ the evaluation of $N_m$ with method B) is unstable because the asymptotic expression of 
$r_3$: $r_3^{asym}$ (see Eq.~\ref{rNasym}) have a couple of zeros in this range (due to the beta coefficients, therefore it is very sensitive to subleading corrections and the truncation), which produces divergences for the theoretical expression 
that we use to determine $N_m$. This is nothing but the reflection of the fact 
that we are in a transition region before we reach the behavior expected for 
$N_l \rightarrow \infty$. In this limit we expect, not only the disappearance of the $u=1/2$ infrared renormalon, but its transformation into a ultraviolet 
 $u=-1/2$ renormalon (so that the perturbative series is sign alternating), for which the normalization can be computed in the large $N_l$ limit 
 \cite{Beneke:1994sw}: 
 $N_m^{(\rm large\; N_l)}=\frac{4}{3}\frac{e^{\frac{5}{6}}}{\pi}=0.976564$. Our evaluation indeed converges towards this value for $N_l>40$.

We have also done some fit-play of the $N_l$ dependence of $N_m$ for small $N_l$ 
using a polynomial function: 
$N_m(N_l)=N_m(0)+d_1N_l+d_2N_l^2+\cdots$. We observe that subsequent coefficients $d_n$ 
get smaller as we increase $n$ for small $N_{l}$, 
with the leading coefficient, $d_1$, of order $\sim -10^{-2}$
 (see also Table \ref{tab:Nm}).

Since we have a reliable determination of $N_m$, we can obtain a prediction for the high order coefficients of the perturbation 
expansions of the pole mass and static potential. We explicitly show them in Table \ref{tab:r3} for the cases of $r_3$ and $r_4$, obtained using Eq.~(\ref{hms}) with 
$\mu=\m_b$ and $h_N(\m_b)=0$. 
The error is fixed by combining in quadrature the error of $N_m$ with the 
error of subleading $1/N$ effects (adding and subtracting the last known term). This last effect is way subleading compared with the uncertainty of $N_m$, which completely dominates the error. Note that the effect of renormalons located at $u=1$ or beyond would produce exponentially suppressed corrections to the $1/N$ expansion. 

\begin{table}
\caption{\label{tab:r3}\it Final predictions for $r_3$ and $r_4$ for different values of $N_l$.}
\begin{tabular}{|c|c|c|c|c|c|c|c|} \hline
$N_f$ &  0 & 1 & 2 & 3 & 4 & 5&6\\ \hline
$4r_3/3$ & 3562(173) & 2887(133) & 2291(98) & 1772(82) & 1324(81) & 945(92)&629(191)
\\ 
\hline
$4r_4/3\times10^{-4}$ & 8.76(42) & 6.66(31) & 4.94(21) & 3.54(16) & 2.44(15) & 1.58(15) &0.95(29)
\\ 
\hline
\end{tabular}
\end{table}

We now compare our predictions with earlier estimates. The quality of large-$\beta_0$ predictions is worse \cite{Beneke:1994qe}. 
This is to be expected as they do not incorporate the right large-$N$ asymptotic behaviour. Dispersion-like analyses \cite{Kataev:2010zh} also seem to have problems to capture the right asymptotic as they yield significantly smaller numbers than the ones obtained here. In 
Ref.~\cite{Pineda:2001zq} (see also Ref.~\cite{Pineda:2002se} for $N_l=0$) the NNLO prediction from method B) was used, which is around one sigma away from our new number. In App.~C of 
Ref.~\cite{Ayala:2012yw}, a variant of this 
method using Pad\'e approximant was worked out and the number obtained was quite similar.
There has also been a recent prediction of $r_3$, 
made in Ref.~\cite{Sumino:2013qqa} by demanding stability of the perturbation
expansion of the heavy quarkonium energy (in the static 
limit) to the 
next order. Our numbers are bigger than his for small $N_l$. Note though that for large $N_l$, we get similar numbers. This points to a different $N_l$ dependence, which in our case is more pronounced. Finally, for $N_l=0$ we can also compare with Eq. (13) of Ref.~\cite{Bali:2013qla}. Their value for $r_3$ is in agreement with ours within one standard deviation. This is quite remarkable as that method is completely different, based on lattice simulations, and, therefore, with different systematics.

\section{Bottom mass from Heavy Quarkonium}
\label{Sec:mbot}
The determination of the pole mass from the $\Upsilon(1S)$ mass 
is plagued by large uncertanties due to the pole mass renormalon. These errors propagate to the determination 
 of the bottom $\MS$ mass $\m$ [$\equiv \m(\m)$].  To avoid this problem we determine the RS bottom
mass $m_{\RS}$ instead. 
 $\m$ can then be obtained from its relation with the $m_{\RS}$ mass. The use of $m_{\RS}$ is 
convenient  because it has no (leading infrared)
renormalon ambiguity, and the renormalon cancellation in the 
quarkonium mass $M_{\Upsilon(1S)}$ is implemented automatically.

\subsection{Renormalon subtracted scheme}
\label{sec:OSvsRSmb}

Formally, the RS mass is defined by subtracting the leading renormalon singularity to the pole mass. For the Borel transform this means 
\begin{equation}
B[m_{\RS}(\nu_f)]\equiv B[m]-N_m \pi \nu_f \frac{1}{(1-2 u)^{1+\nu}}\left(1+{\widetilde c}_1(1-2 u)+ 
{\widetilde c}_2(1-2 u)^2+\ldots \right) \ ,
\label{mrsdef}
\end{equation}
where $m$ is the pole mass ($m_q$), and 
we use the notations of Eqs.~(\ref{tr})-(\ref{BSrenan}).
Therefore, we have the following explicit expression for $m_{\RS}$:
\begin{equation}
m_{\RS}(\nu_f)=m -
\delta m_{\RS}\label{mrs1}
\,,
\end{equation} 
where $\delta m_{\RS}$ is the residual mass (we recall that ${\widetilde c}_0=1$):
\begin{equation}
\delta m_{\RS}(\nu_f)
=
N_m \pi \nu_f \sum_{N=0}^{\infty} (2\beta_0)^N a_{-}^{N+1}(\nu_f) \sum_{n=0}^\infty {\widetilde c}_n\frac{\Gamma(\nu+N+1-n)}{\Gamma(\nu+1-n)} \ .
\label{deltamrs}
\end{equation} 
Note that we work in the theory with three active flavours only, as the charm decouples at large orders in perturbation theory (which is the regime $\delta m_{\RS}$ deals with).

Equation (\ref{mrs1}) is still formal. In practice, one rewrites $m$ in terms of $\m$ using Eq. (\ref{mqbmq}) and reexpands the perturbation series in Eq. (\ref{deltamrs}) around the same 
coupling $a_{-}(\mu)$, 
at fixed but otherwise arbitrary scale $\mu$:
\bes
\label{mrs2}
\bea
m_{\RS}(\nu_f) &=&
\m \left[ 1 +\sum_{N=0}^{\infty} h_N(\nu_f) a_{-}^{N+1}(\nu_f) \right] 
\label{mrs2a}
\\
\Rightarrow \;\;\;
m_{\RS}(\nu_f)&=&\m \left[ 1 +\sum_{N=0}^{\infty} {\widetilde h}_N(\nu_f;\mu) a_{-}^{N+1}(\mu) \right] \ ;
\label{mrs2b}
\eea
\ees
where $h_N(\nu_f)$ is determined from Eq.~(\ref{hms})
(with $\mu=\nu_f$ and with the sum truncated at $s=3$) for $N=0,1,2$. For $N \geq 3$ we take $h_N(\m_b)=0$. The coefficients 
${\widetilde h}_N(\nu_f;\mu)$
in Eq.~(\ref{mrs2b}) are obtained by expanding $a_{-}(\nu_f)$ in
the expansion (\ref{mrs2a}) in powers of $a_{-}(\mu)$. 
 This procedure ensures that the renormalon behaviour is cancelled order by order 
in $a_{-}(\mu)$. 
Note that $m_{\RS}(\nu_f)$ does not depend on $\mu$ (it will, but only marginally, 
 when we truncate the infinite sum in Eq.~(\ref{mrs2})). On the other hand the coefficients $h_N$ are functions of $\nu_f$, $\mu$, and $\m$, and are much smaller than $r_N(\mu)$.

Another possibility is to define a modified renormalon-subtracted 
 (RS') mass $m_{\rm RS'}(\mu)$, Ref.~\cite{Pineda:2001zq},
where subtractions start at the level $\sim a^2$ [i.e., $N=1$
in Eq.~(\ref{mrs1})]
\begin{equation}
m_{\RS'}(\nu_f)=m-N_m \pi \nu_f \sum_{N=1}^{\infty} (2\beta_0)^N a_{-}^{N+1}(\nu_f) 
\sum_{s=0}^\infty {\widetilde c}_s\frac{\Gamma(\nu+N+1-s)}{\Gamma(\nu+1-s)} \ ,
\label{mrsp1}
\end{equation}
and this leads to a relation analogous to Eqs.~(\ref{mrs2})
\bes
\label{mrsp2}
\bea
m_{\RS'}(\nu_f) &=&\m \left[ 1 + \frac{4}{3} a_{-}(\nu_f) + 
\sum_{N=1}^{\infty} h_N(\nu_f) a_{-}^{N+1}(\nu_f) \right] 
\label{mrsp2a}
\\
\Rightarrow \;\;\;
m_{\RS'}(\nu_f)&=&\m \left[ 1 + \frac{4}{3} a_{-}(\mu) + 
\sum_{N=1}^{\infty} {\widetilde h}_N^{\prime}(\nu_f;\mu) a_{-}^{N+1}(\mu) \right] \ ,
\label{mrsp2b}
\eea
\ees
where ${\widetilde h}_N^{\prime}(\nu_f;\mu)$ in Eq.~(\ref{mrsp2b}) are obtained by expanding $a_{-}(\nu_f)$ in Eq.~(\ref{mrsp2a}) in powers of $a_{-}(\mu)$.
The explicit relation between the two Borel transforms is
\begin{equation}
B[m_{\RS'}(\nu_f)]\equiv B[m_{\RS}(\nu_f)]+N_m \pi \nu_f  \left(1+{\widetilde c}_1+{\widetilde c}_2+\ldots \right) \ .
\label{mrspmrs}
\end{equation}

\subsection{$\Upsilon(1S)$ mass}
\label{sec:Upsilonmass}

The perturbation expansion of the $\Upsilon(1S)$ mass  
is presently known up to ${\cal O} (m_b a^5)$
\bea
M^{(th)}_{\Upsilon(1S)}& = &2m_b - \frac{4 \pi^2}{9} m_b a_-^2(\mu) 
{\Big \{} 1 + a_-(\mu) \left[ K_{1,0} + K_{1,1} L_p(\mu) \right] +
a_-^2(\mu) \sum_{j=0}^2 K_{2,j} L_p(\mu)^j
\nonumber\\
&& + a_-^3(\mu) 
\left[ 
K_{3,0,0} + K_{3,0,1} \ln a_-(\mu)+ \sum_{j=1}^3 K_{3,j} L_p(\mu)^j 
\right]
+ {\cal O}(a_-^4) {\Big \}} \ ,
\label{Ebb}
\eea
where $\mu$ is the renormalization scale,
$m_b$ is the pole mass of the bottom quark and 
\bea
L_p(\mu) & = & \ln \left( \frac{\mu}{(4 \pi/3) m_b 
 a_{-}(\mu)
} \right) \ ,
\label{Lp}
\eea
The numerical expressions of the coefficients $K_{i,j}(N_f)$ and
$K_{3,0,j}$ are given for reference in Appendix \ref{app2}.

As we have already discussed throughout the paper, 
it is compulsory to implement the cancellation of the leading 
infrared renormalon ($u=1/2$) in the above 
perturbation series to get a convergent series. 
We do so by working in the RS scheme. In practice this means to rewrite $m_b$ in terms of 
$m_{b,\RS}$ in Eq.~(\ref{Ebb}). The resulting expression reads
$M_{\Upsilon(1S)}$:
\bea
\frac{M^{(th)}_{\Upsilon(1S)}}{ m_{b,\RS}(\nu_f)} & = & 2
+ \left[ 
2 \pi N_m b a {\cal K}_0 - \frac{4 \pi^2}{9} a^2 
\right]
+ \left[
2 \pi N_m b a^2 \left( {\cal K}_1 + z_1 {\cal K}_0 \right)  
- \frac{4 \pi^2}{9} a^3 \left( K_{1,0} + K_{1,1}L_{\RS}
\right) \right]
\nonumber\\ &&
+ \left[
2 \pi N_m b a^3 \left( {\cal K}_2 + 2 z_1 {\cal K}_1 + z_2 {\cal K}_0 \right) 
- \frac{4 \pi^2}{9} \left( a^4 \sum_{j=0}^2 K_{2,j} L_{\RS}^j
+ b a^3 \pi N_m {\cal K}_0
\right) \right]
\nonumber\\ &&
\nn
+ {\Bigg[}
2 \pi N_m b a^4 \left( {\cal K}_3 + 3 z_1 {\cal K}_2 + (2 z_2+z_1^2) {\cal K}_1
+ z_3 {\cal K}_0 \right)
\\&&
- \frac{4 \pi^2}{9} {\bigg [} 
a^5 \left( K_{3,0,0} + K_{3,0,1} \ln a+ \sum_{j=1}^3 K_{3,j} L_{\RS}^j \right)
\nonumber\\ &&
+ b a^4 \pi N_m \left( 
K_{1,0} {\cal K}_0 + (L_{\RS}-1) K_{1,1} {\cal K}_0
+ {\cal K}_1 + z_1 {\cal K}_0 \right) {\bigg ]}
{\Bigg]} \ ,
\label{MUs2}
\eea
where we denoted 
\begin{subequations}
\label{ambLNKN}
\bea
a & \equiv & a_{-}(\mu) = a(\mu,N_f=3) \ , \quad b  \equiv  b(\nu_f) = \frac{\nu_f}{m_{b,\RS}(\nu_f)} \ , \quad N_m  =  N_m(N_l=3) \ ,
\label{am}
\\
L_{\RS} &\equiv& L_{\RS}(\mu) =
\ln \left( \frac{\mu}{ (4 \pi/3) m_{b,\RS}(\nu_f) a_{-}(\mu) } \right) \ ,
\quad 
{\cal K}_N =  (2 \beta_0)^N  \left[ 1 + \sum_{s=1}^{3} 
{\widetilde c}_s \frac{\Gamma(\nu+N+1-s)}{\Gamma(\nu+1-s)} \right]  \ .
\nn
\\
\label{KN}
\eea
\end{subequations}
In the expression (\ref{MUs2}) for $M_{\Upsilon(1S)}$, the terms of the same order 
$(\nu_f/\m_b) a^n$ and $a^{n+1}$ were combined in common brackets $[ \ldots ]$, in order to
account for the renormalon cancellation. 

If using the RS' mass in our approach  instead, the above expressions
are valid without changes, except that $m_{\RS} \mapsto m_{\RS'}$ and
${\cal K}_0 \mapsto 0$ (and: $h_0(\mu) \mapsto 4/3$).

We note that we take $N_l=3$ active flavours and that the expression above does not incorporate yet the charm quark effects. The leading one is due 
to the potential Eq.~(\ref{eq:vpmc2}) and reads (see, for instance, \cite{Eiras:2000rh})
\be
\label{eq:Mch}
\delta M_{\Upsilon(1S)}=
\frac{\m_b 4(\pi a_-)^2}{9}\frac{a_-}{3}
\left(
\frac{11}{3}-\frac{3}{2}\pi {\bar \rho}+4{\bar \rho}^2-2\pi{\bar \rho}^3
+\frac{-4+2{\bar \rho}^2+8{\bar \rho}^4}{\sqrt{{\bar \rho}^2-1}}{\rm ArcTan}\left(\sqrt{\frac{{\bar \rho}-1}{{\bar \rho}+1}}\right)
\right)
\,,
\ee
where ${\bar \rho}=3\m_c/(2\m_b\pi a_-)$. 
It produces a shift of order 1 MeV, completely negligible in comparison with other uncertainties. 
This has also been stressed in Ref.~\cite{Brambilla:2001qk}. Nevertheless, it got obscured because specific numbers were given with 
$N_f=4$.

\subsection{Bottom mass determination}
\label{mb}

$m_{\RS}(\nu_f)$ is determined from the
condition (see Eq. (\ref{MUs2}))
\be
M^{(th)}_{\Upsilon(1S)}
= M^{(exp)}_{\Upsilon(1S)} \; ( = 9.460 \ {\rm GeV} ) \ .
\label{MUexp}
\ee

We now investigate the dependence of our results on the 
theoretical and experimental parameters. 
To estimate the errors, we vary 
$\mu$, $\nu_f$, $\als$ and
$N_m$ as follows: $\mu=2.5^{+1.5}_{-1}$ GeV, $\nu_f=2\pm 1$ GeV,
  $\als(M_z)=0.1184(7)$ \cite{PDG2012} (with decoupling at 4.2 GeV and 1.27 GeV for the bottom 
and charm $\MSbar$ masses,
respectively) and $N_m=0.563(26)$. For the RS
scheme, we obtain the following result\footnote{Here and in the following,
in the determination of $m_{\MS}(m_{\MS}) \equiv \m(\m) \equiv \m$, 
we have used our estimate of the four-loop relation.}\footnote{Note that the scale dependence of $\m$ 
is the one associated to the fit to $m_{\RS}$.}
\bea
\label{MRSdet}
m_{b,\RS}(2\;{\rm GeV})
&=&
4\,437^{-11}_{+43}(\mu)^{-3}_{+5}(\nu_f)^{-2}_{+2}(\als)^{-41}_{+41}(N_m)\;{\rm MeV};
\\
\label{MMSRSdet}
 \m_{b}(\m_{b})
&=&4\,201^{-10}_{+39}(\mu)^{-3}_{+5}(\nu_f)^{-6}_{+6}(\als)^{-17}_{+17}(N_m)\;{\rm MeV}.
\eea
For the RS' scheme, we obtain the result (with the same variation of the
parameters)
 \bea
\label{MRSprimedet}
m_{b,\RS'}(2\;{\rm GeV})&=&4\,761^{-16}_{+41}(\mu)^{-3}_{+5}(\nu_f)
^{+4}_{-3}(\als)^{-26}_{+26}(N_m)\;{\rm MeV};
\\
\label{MMSRSprimedet}
\m_{b}(\m_{b})
&=&4\,206^{-14}_{+36}(\mu)^{-2}_{+4}(\nu_f)^{-5}_{+5}(\als)^{-17}_{+17}(N_m)\;{\rm MeV}.
\eea

%%%%Figure figM1SRS%%%%
\begin{figure}[h]
\hspace{-0.1in}
\epsfxsize=4.8in
\centering\includegraphics[width=140mm]{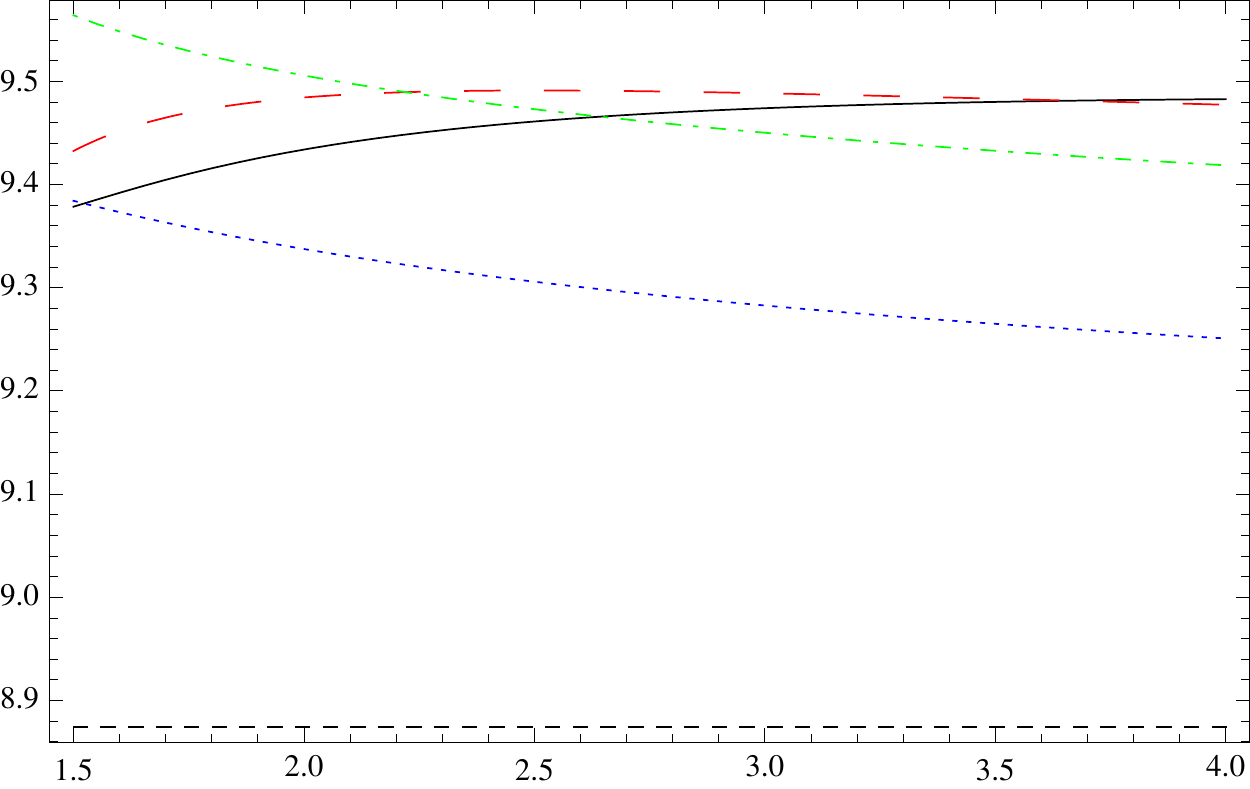}
%\centerline{\epsffile{MUpsilonRS.eps}}
\caption{{\it We 
plot $2 m_{b,\RS}(2\;{\rm GeV})$ (dotted line), and the LO (short-dashed line), NLO (dot-dashed line), NNLO
  (long-dashed line) and NNNLO
  (solid line) predictions for the
  $\Upsilon(1S)$ mass in terms of $\mu$ in the {\rm RS} scheme. 
The value of $m_{b,\RS}(2\;{\rm GeV})$ is taken from Eq. (\ref{MRSdet}).
All the scales are in GeV.}}
\label{figM1SRS}
\end{figure}
%%%%End Figure figM1SRS%%%%
%%%%Figure figM1SRSprime%%%%
\begin{figure}[h]
\hspace{-0.1in}
\epsfxsize=4.8in
\centering\includegraphics[width=140mm]{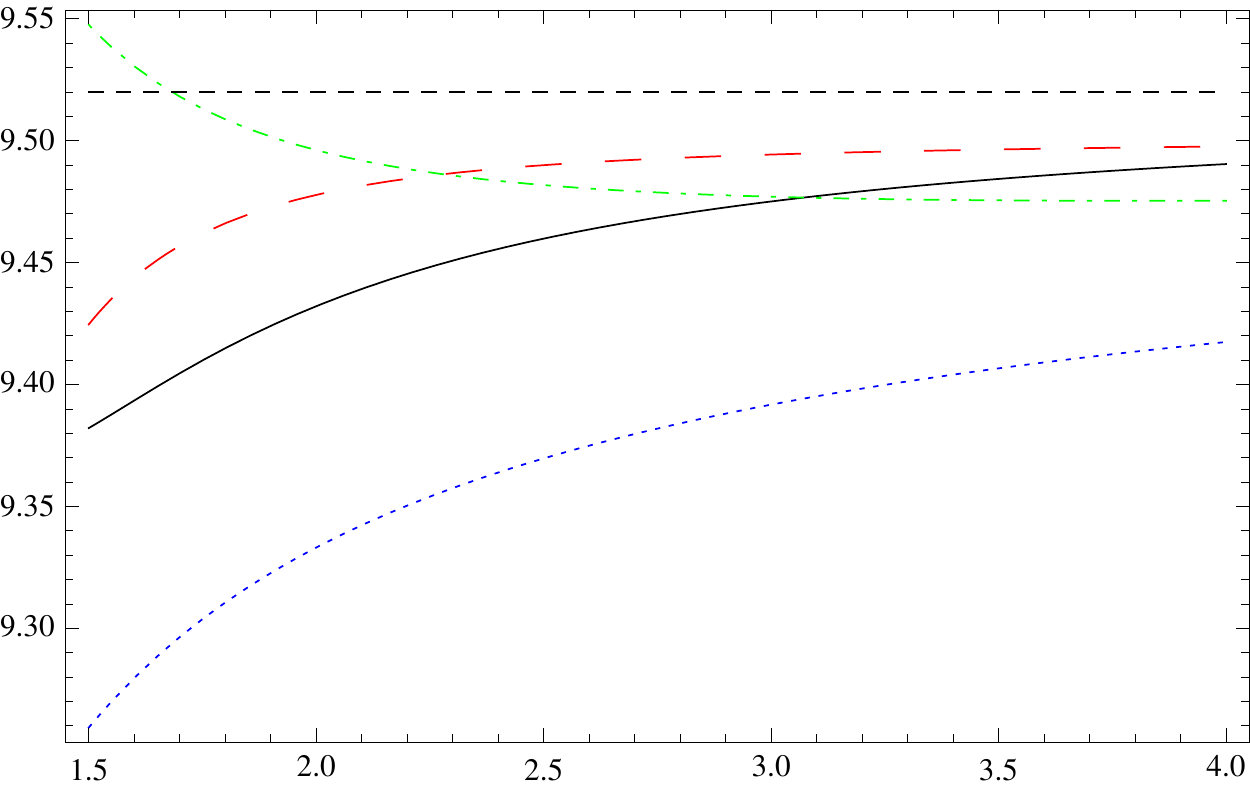}
%\centerline{\epsffile{MUpsilonRSprime.eps}}
\caption{{\it 
 As in Fig.~\ref{figM1SRS}, but now with $2m_{b,\RS'}(2\;{\rm GeV})$ and the corresponding predictions for the  $\Upsilon(1S)$ mass in the RS' scheme. The value of $m_{b,\RS'}(2\;{\rm GeV})$ is taken from Eq.~(\ref{MRSprimedet}).
}}
\label{figM1SRSprime}
\end{figure}
%%%%End Figure figM1SRSprime%%%%

For the central values obtained in Eqs.~(\ref{MRSdet})-(\ref{MMSRSprimedet}), we can visualize the relative size of the different terms of 
the perturbative expansion of $M_{\Upsilon(1S)}$ and $m_{\RS(\RS')}$. In the RS we obtain (for both expansions we take $\mu=2.5$ GeV)
\bea
M_{\Upsilon(1S)}&=& 
 ( 
8875+431+166+18-30
 ) \ {\rm MeV} \,, 
\\
m_{\RS}(2\;{\rm GeV})&=&
 ( 
4201+189+36+12-0
 ) \ {\rm MeV} \ ,
\eea
where the -0 of the last equality is accidental for the specific scale chosen (see Fig.~\ref{figM1SRS}).

In the RS' we obtain
\bea
M_{\Upsilon(1S)}&=&
 ( 
9521-150+112+8-31
 ) \ {\rm MeV} \,, 
\\
m_{\RS'}(2\;{\rm GeV})&=&
 ( 
4206+476+60+18+1
 ) \ {\rm MeV} \ . 
\eea
We observe a nicely convergent perturbative series in the relation between the RS(RS') masses and the MS mass. In the perturbative relation 
between the $\Upsilon(1S)$ mass and the RS(RS') masses we also observe a convergent series except when we consider the difference between the NNLO and NNNLO results. 
In Figs.~\ref{figM1SRS} and \ref{figM1SRSprime}, we plot the scale dependence of the LO, NLO, NNLO and NNNLO predictions for
the $\Upsilon(1S)$ mass in the RS and RS' scheme in order to observe the pattern of convergence for different values of $\mu$, as well as to show 
the scale dependence of our results. 
We observe a convergent series except when we consider the difference between the NNLO and NNNLO results. 
The latter shows a stronger scale dependence. This could be expected, as at this order the hard and ultrasoft scales enter into play. 
The hard scale enters through the Wilson coefficients of the NRQCD Lagrangian (see, for instance, Sec.~4 in Ref.~\cite{BPSV}). Specially problematic 
is the appearance of the ultrasoft scale, since it is potentially a rather small scale. Whether this scale can be treated within perturbation 
theory can
only be elucidated by higher order computations, as well as by analyses using renormalization group techniques. Without such analyses it 
is not possible to unambiguously set the scale of the 
NNNLO contribution, and any estimate of higher order effects due to the ultrasoft contributions will suffer from some scheme dependence. 
Nevertheless, we can observe some general trends. The ultrasoft logarithmic dominated terms \cite{Kniehl:1999ud,BPSV} yield a positive 
contribution to the NNNLO term. This contribution would be even bigger if we set the scale of one of the powers of $\als$ at the ultrasoft 
scale (as the effective field theory suggests). This would go in the direction of making the NNNLO result (and also the value of the bottom 
mass) smaller and improve convergence. Nevertheless, without a renormalization group analysis we can not make this discussion 
more quantitative. Finally,  
without a better control of the ultrasoft effects it would be premature to consider 
nonperturbative corrections, which we neglect in this analysis. As for the large-$\beta_0$ approximation we observe that they give numbers in the right ballpark, yet one should keep in mind that this comparison will depend on the constant term in the ultrasoft logarithm.\footnote{We also remind that the large-$\beta_0$ approximation does not have the right asymptotics. Therefore, any eventual agreement will deteriorate at higher orders.}

To this analysis one has to add charm effects. The leading correction to the heavy quarkonium spectrum can be found in Eq. (\ref{eq:Mch}). 
It produces a negligible correction $\sim -0.6$ MeV. Therefore, it barely changes our determination above. 
The corrections to the relation between 
$m_{\RS}$ and $\m$, Eqs.~(\ref{SmNl0})-(\ref{delmc}) are more important, though still quite small (at the MeV level).
The correction that we find (from the ${\cal O}(a^2)$ and ${\cal O}(a^3)$ terms) is $\sim -2.4+0.4\simeq -2$ MeV 
(for our standard value $\mu=2.5$ GeV). 
We introduce this shift in our final number for the bottom mass, which reads
\be
\m(\m) = 
4.201(43)\; {\rm GeV}
\,, 
\ee 
where we have done the average of the RS and RS' determination, rounding the $\pm$ variation of each parameter to the maximum, and added the errors in quadrature. Let us also observe that the difference between the NNLO ($m_{\RS}=4.421$ GeV and $\m=4.187$) 
and NNNLO ($m_{\RS}=4.437$ GeV and $\m=4.201$ GeV) evaluation is inside the range of scale variation we consider, which gives extra confidence in our error analysis. It is also comforting that the inclusion of the NNNLO correction (and the charm effects) do not shift much the central value with respect the NNLO evaluation made in Ref.~\cite{Pineda:2001zq} (or with $m_{\RS}=4.425$ GeV and $\m=4.211$ GeV, which 
is slightly different than the result obtained in \cite{Pineda:2001zq}, due 
to the different values of the parameters), which corresponds to working with 
$N_f=4$ active massless quarks. If we do the NNNLO determination also with $N_f=4$ we obtain $m_{\RS}=4.482$ GeV and $\m=4.261$ GeV. 
To be compared with the same analysis with $N_f=3$. The $N_f=3$ case shows a better convergence, as expected.
 
\section{Conclusions}
\label{concl}

In this paper we have considered different improvements over the analysis made in Ref.~\cite{Pineda:2001zq}.

First, we have studied whether (or when) the charm quark decouples in the perturbative relation between the pole and the $\MS$ mass. 
For the case of the heavy quarkonium mass (versus the pole mass) it was seen that the charm quark decoupled and 
it was a good approximation to consider the theory with only three active massless flavours \cite{Brambilla:2001qk}. 
In this paper we have concluded that one should also use this approximation for the relation between the pole and the $\MS$ mass. This leads to shifts of order of 1 MeV in both relations making them negligible in comparison with other uncertainties. 

Second, we have obtained an improved determination of the normalization of the leading pole mass (and static potential) 
renormalon. For $N_l=0$ and $N_l=3$, they read\footnote{As they suffer from different systematics, we can consider combining 
this result with $N_m\Big|_{N_l=0}= 0.620(35)$  \cite{Bali:2013qla} and obtain an even more accurate value:  $N_m\Big|_{N_l=0}= 0.608(22)$.}
\be
\label{eq:Nmfinal}
N_m\Big|_{N_l=0}= 0.600(29) \,, \qquad  N_m\Big|_{N_l=3}= 0.563(26)
\,.
\ee 
This has an immediate impact in 
the determination of the heavy quark mass from the heavy quarkonium ground state mass but it is also applicable to other observables 
in heavy quark physics. This improvement is twofold. On the one hand the existence of the three-loop expression of the static potential allows us to determine the normalization to one order higher in the corresponding expansion. On the other hand, we obtain the normalization 
directly from the last known coefficient of the perturbation expansion. This leads to a more stable result 
compared with previous approaches, as it has already been observed in lattice simulations \cite{Bali:2013pla} 
for the case of the self-energy of a static quark.

Finally, we have included the complete NNNLO correction to the perturbative expression of the 
$\Upsilon(1S)$ mass and determined the bottom quark mass:
\be
\m(\m) = 
4.201(43)\; {\rm GeV}
\,,
\ee 
using the renormalon subraction scheme. 
In this analysis we have worked with three active flavours, as motivated by the previous discussion. We have also used the updated 
value of $N_m$ obtained in Eq. (\ref{eq:Nmfinal}).  By including the complete NNNLO expression we can study this term without 
scheme ambiguity. At this order ultrasoft effects appear for the first time. Consistent with this fact we observe that 
the NNNLO result is more scale dependent than the NNLO one, and the convergence of the perturbative series deteriorates somehow.
It remains to be analyzed whether a renormalization group analysis could reduce the scale dependence of the result and improve the convergence. Yet, the magnitude of the NNNLO correction is small, so that the final number is quite close to the value obtained by the NNLO analysis made in Ref.~\cite{Pineda:2001zq}. On the other hand, our determination is considerably smaller than the NNNLO determinations in Refs.~\cite{Penin:2002zv,Ayala:2012yw}. \cite{Penin:2002zv} worked with $N_f=4$ and an estimate for $a_3$ but did not implement explicitly the cancellation of the renormalon. 
In Ref.~\cite{Ayala:2012yw} the $\MSbar$ scheme and another related scheme were used, the calculation was performed at NNNLO for the $\Upsilon(1S)$ mass with $N_f=3$ but $N_f=4$ was used in the relation for $m/\m$, and a lower soft renormalization scale $\mu \approx 2$ GeV was used; the (strong) $m_c$ effects were not under control at NNNLO (due to the use of $N_f=4$), besides producing a small mismatch in the renormalon cancellation. Our number is somewhat in the middle between two recent determinations from bottomonium NR sum rules \cite{Hoang:2012us,Penin:2014zaa}. Both of them worked with $N_f=4$ active massless quarks, though the latter reference estimated the shift produced by the finite mass charm quark effects. 
\cite{Hoang:2012us} performed a partial NNLL computation. The difference with the (also partial) NNLL determination in 
Ref.~\cite{Pineda:2006gx} stems from extra/different terms incorporated in the analysis.  \cite{Penin:2014zaa} performs a partial NNNLO 
computation. Note also that our number is similar to the number obtained from a lattice HQET determination 
\cite{Bernardoni:2013xba}, and not very far away from the numbers 
obtained using low-$n$ or finite-energy bottomonium sum rules \cite{Kuhn:2007vp,Bodenstein:2011fv}, or a lattice determination using NRQCD \cite{Lee:2013mla}. Those have complete different systematics. This agreement may indicate that nonperturbative corrections are indeed small, as advocated in 
Refs.~\cite{Brambilla:2001fw,Brambilla:2001qk}.

\begin{acknowledgments}
\noindent
We thank M.~Steinhauser for bringing us \cite{Bekavac:2007tk} to our attention. 
This work was supported in part by FONDECYT (Chile) Grant No.~1130599, 
DGIP (UTFSM) internal project USM No.~11.13.12,  the Spanish grants FPA2010-16963 and FPA2011-25948, and the Catalan grant
SGR2014-1450.
\end{acknowledgments}

\appendix

\section{Relations between different couplings}
\label{app3}

The relation between $a_{+}(\nu_f)$ and $a \equiv a_{-}(\mu)$ can be written in the form
\bea
a_{+}(\nu_f) & = & a \left[ 1 + z_1 a + z_2 a^2 + z_3 a^3
+ {\cal O}(a^4) \right] \ ,
\label{apm}
\eea
where the coefficients $z_1$ account for the $N_f=4 \mapsto 3$ quark threshold
effects and the (subsequent) renormalization group running from $\nu_f$ to $\mu$. The
threshold effects are taken at the three loop level according to Ref.~(\cite{CKS}),
and the renormalization group running at the four loop level. The resulting coefficients $z_j$ are:
\bea
z_1 & = & x_1 + y_1 \ , \qquad z_2 = x_2 + 2 x_1 y_1 + y_2 \ ,
\nonumber\\
z_3 & = & x_3 + 3 x_2 y_1 + x_1 y_1^2 + 2 x_1 y_2 + y_3 \ .
\label{zjs}
\eea
Here, the coefficients $x_j$ reflect the three-loop quark threshold matching
for $N_f=4 \mapsto 3$ at the chosen threshold scale $\nu_f$,
\be
x_1 = - k_1 \ , \quad 
x_2 = - k_2 + 2 k_1^2 \ , \quad
x_3 = - k_3 + 5 k_1 k_2 - 5 k_1^3 \ ,
\label{xjs}
\ee
where the expressions for $k_j$  ($j=1,2,3$) 
are given in Ref.~\cite{CKS} ($k_1 =  -\ell_h/6$, etc.),
with the logarithm there being $\ell_h = \ln(\nu_f^2/\m_c^2)$
and $N_{\ell}=3$ (see also Appendix D of Ref.~\cite{Ayala:2012yw}).
The coefficients $y_j$ reflect the (subsequent) renormalization group running from 
$\nu_f$ to $\mu$ (with $N_f=3$)
\be
y_1 =  \beta_0 \ln \left( \frac{\mu^2}{\nu_f^2} \right) \ , \quad
y_2 = y_1^2 + c_1 y_1 \ , \quad
y_3 = y_1^3 + \frac{5}{2} c_1 y_1^2 + c_2 y_1 \ .
\label{yjs}
\ee
Here, $c_j \equiv \beta_j/\beta_0$.

\section{Numerical values of the coefficients of the binding energy}
\label{app2}

In this Appendix we summarize, for reference, 
the numerical values of the coefficients
$K_{i,j}(N_l)$ and $K_{3,0,j}$ entering the perturbation
expansion of the binding energy $E_{q \bar q}$, Eq.~(\ref{Ebb}). 
Note that $N_l=3$ for bottomonium. Recall also that, in our convention, $\beta_0=(1/4)(11 -2 N_l/3)$ and 
$\beta_1=(1/16)(102-38 N_l/3)$.
These expressions can be extracted from the results obtained
or given in
Refs.~\cite{Fischler:1977yf,Billoire:1979ih,Schroder:1998vy,Pineda:1997hz,BPSV,Kniehl:2002br,Penin:2002zv,Smirnov:2008pn,Anzai:2009tm,Smirnov:2009fh}:
\bea
K_{1,0}(N_l) & = & \frac{1}{18} (291 - 22 N_l) = 16.1667 - 1.22222 N_l \ , 
\quad
K_{1,1}(N_l)= 4 \beta_0 \ ;
\label{K1j}
\eea
\begin{subequations}
\bea
K_{2,0}(N_l) & = & 337.947 - 40.9649 N_l + 1.16286 N_l^2 \ ,
\nonumber\\
K_{2,1}(N_l) & = & 231.75 - 32.1667 N_l + N_l^2 \ ,
\nonumber\\
K_{2,2}(N_l) & = & 12 \beta_0^2 \ ; 
\label{K2j}
\eea
\end{subequations}
\begin{subequations}
\bea
K_{3,0,0}(N_l) & = & 8041.49 - 1318.36 N_l + 75.263 N_l^2 - 1.25761 N_l^3 \ ,
\nonumber\\
K_{3,0,1}(N_l) & = & \frac{865 \pi^2}{18} =  474.289 \ ,
\label{K30j}
\eea
\end{subequations}
\begin{subequations}
\bea
K_{3,1}(N_l) & = & 6727.62 - 1212.76 N_l + 69.1066 N_l^2 - 1.21714 N_l^3 \ ,
\nonumber\\
K_{3,2}(N_l) & = & 2260.5 - 456.458 N_l + 28.5278 N_l^2 - 0.555556 N_l^3 \ ,
\nonumber\\
K_{3,3}(N_l) & = & 32 \beta_0^3 \ .
\label{K3j}
\eea
\end{subequations}

\end{document}